\documentclass[journal]{IEEEtran}
\pdfoutput=1

\usepackage{algorithm}
\usepackage{algorithm,algpseudocode}

\usepackage{amsmath}    
\usepackage{amssymb}
\usepackage{graphicx}   
\usepackage{multirow}  
\usepackage[scientific-notation=true]{siunitx}
\usepackage{verbatim}   
\usepackage{color}      
\usepackage{subfig}  
\usepackage[draft]{hyperref}   
\usepackage{mathtools}
\usepackage[normalem]{ulem} 
\newcommand{\stkout}[1]{\ifmmode\text{\sout{\ensuremath{#1}}}\else\sout{#1}\fi}
\usepackage{soul} 
\usepackage{kantlipsum}
\usepackage{comment}
\allowdisplaybreaks

\definecolor{ao(english)}{rgb}{0.0, 0.5, 0.0}
\definecolor{dartmouthgreen}{rgb}{0.05, 0.5, 0.06}
\definecolor{deepmagenta}{rgb}{0.8, 0.0, 0.8}
\definecolor{lavenderindigo}{rgb}{0.58, 0.34, 0.92}

\def \asdILP {\delta_{sd}^{\OrigConfigIdx}}

\def \asdPP {\delta_{sd}}
\def \avi {a_{vi}}
\def \bil  {b_{i \ell}}
\def \CAP {\textsc{cap}}
\def \OrigConfigSet {\hat{\Gamma}}
\def \OrigConfigIdx {\hat{\gamma}}
\def \ConfigSet {\Gamma}
\def \ConfigIdx {\gamma}
\def \cost {\textsc{cost}}

\def \deltafic {\delta_{fi}^{\SFCidx}}

\def \Tfic {\textsc{T}_{fi}^{\SFCidx}}

\def \ncopyc {I_c}
\def \ncore {n^{\textsc{core}}}
\def \Ncore {\textsc{N}^{\textsc{core}}}
\def \ncoref {n^{\textsc{core}}_{f}}
\def \lambdaf    {n^{\textsc{core}}_{f}}
\def \ncorefi      {n^{\textsc{core}}_{f_i}}
\def \nlocationf {R_f}
\def \NFV         {\textsc{nfv}}
\def \NFVI {\textsc{nfv}}

\def \redcost {\textsc{red\_cost}}
\def \SD {\mathcal{SD}}
\def \ServiceChainSet {C}
\def \SFCidx {c}
\def \SFCset {C}
\def \ServiceChainIdx {c}
\def \xvic {x_{v}^{\SFCidx i}}
\def \csdvi {p^{sd}_{v,i}} 
\def \csdvsrcone {p^{sd}_{v_s,1}}
\def \csdvone {p^{sd}_{v,1}}
\def \csdvdestlast {p^{sd}_{v_d,n_c}}
\def \csdvlast {p^{sd}_{v,n_c}}
\def \xone {x_{v_s}^{\SFCidx, 1}}
\def \xvone {x_{v}^{\SFCidx, 1}}

\def \qsdli {q^{sd}_{i \ell}} 
\def \zILP {z_{\OrigConfigIdx}}
\def \zRMP {z_{\ConfigIdx}}
\def \myspace {\hspace*{-1.cm} }

\def \SCCopies {Instances}

\def \SCcopies {instances}
\def \SCcopy {instance}

\begin{document}
%
\title{A Scalable Approach for Service Chain (SC) Mapping with Multiple SC {\SCCopies} in a Wide-Area Network \\
		\large {\color{green}This is a preprint electronic version of the article 
		submitted to IEEE JSAC Series on Network Softwarization And Enablers}}
%
%
%

\author{Abhishek~Gupta,~Brigitte~Jaumard,~Massimo~Tornatore,
        and~Biswanath~Mukherjee
\IEEEcompsocitemizethanks{\IEEEcompsocthanksitem A. Gupta, M. Tornatore, and B. Mukherjee
 are with the University of California, Davis, USA.
E-mail: {\{abgupta, mtornatore, bmukherjee}\}@ucdavis.edu; B. Jaumard is with Concordia University, Canada.
E-mail: bjaumard@cse.concordia.ca; M. Tornatore is also with Politecnico di Milano, Italy.
E-mail: massimo.tornatore@polimi.it}
\thanks{}}

\markboth{Journal of \LaTeX\ Class Files,~Vol.~14, No.~8, August~2015}%
{Shell \MakeLowercase{\textit{et al.}}: Bare Demo of IEEEtran.cls for IEEE Journals}
%

\maketitle

\begin{abstract}
Network Function Virtualization (NFV) aims to simplify deployment of network services by running Virtual Network Functions (VNFs) on commercial off-the-shelf servers. Service deployment involves placement of VNFs and in-sequence routing of traffic flows through VNFs comprising a Service Chain (SC). The joint VNF placement and traffic routing is called SC mapping. 
In a Wide-Area Network (WAN), a situation may arise where several traffic flows, generated by many distributed node pairs, require the same SC; then, a single {\SCcopy} (or occurrence) of that SC might not be enough. SC mapping with multiple SC {\SCcopies} for the same SC turns out to be a very complex problem, since the sequential traversal of VNFs has to be maintained while accounting for traffic flows in various directions.  

Our study is the first to deal with the problem of SC mapping with multiple SC {\SCcopies} to minimize network resource consumption. We first propose an Integer Linear Program (ILP) to solve this problem. Since ILP does not scale to large networks, we develop a column-generation-based ILP (CG-ILP) model. However, we find that exact mathematical modeling of the problem results in quadratic constraints in our CG-ILP. The quadratic constraints are made linear but even the scalability of CG-ILP is limited. Hence, we also propose a two-phase column-generation-based approach to get results over large network topologies within reasonable computational times. Using such an approach, we observe that an appropriate choice of only a small set of SC {\SCcopies} can lead to a solution very close to the  minimum bandwidth consumption. Further, this approach also helps us to analyze the effects of number of VNF replicas and number of NFV nodes on bandwidth consumption when deploying these minimum number of SC instances. 
\end{abstract}


%
\IEEEpeerreviewmaketitle

%
%
%
%
\section{Introduction}
\label{intro}
\IEEEPARstart{T}{}raditionally, communication networks have deployed network services through proprietary hardware appliances (e.g., network functions such as  firewalls, NAT, etc.) which are statically configured. With rapid evolution of applications, networks require agile and scalable service deployment. 

Network Function Virtualization (NFV) \cite{etsi_nfv} offers a solution for an agile service deployment. NFV envisions traditional hardware functionality as software modules called Virtual Network Functions (VNFs). VNFs can be run on commercial-off-the-shelf hardware such as servers and switches in datacenters (DCs), making service deployment agile and scalable. 

\begin{figure*}
  \centering
  \begin{tabular}{cc}
     \subfloat[][One single SC {\SCcopy} mapping: requests $r_3$ and $r_2$ take longer paths. All traffic requests get mapped to the single instance.]{\label{fig:a}\includegraphics[width=.55\textwidth, scale=1]{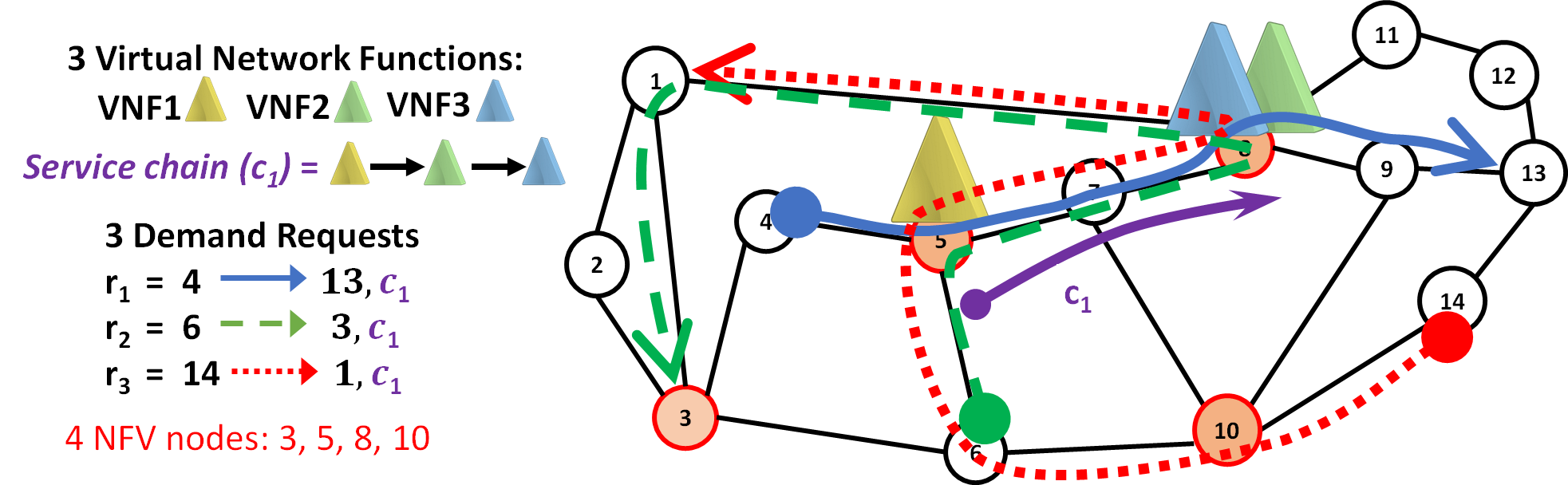}}
    & \subfloat[][Two SC {\SCcopies} mapping: requests $r_3$ and $r_2$ take shorter paths. Here, each traffic request chooses to map to one of two given instances.]{\label{fig:b}\includegraphics[width=.36\textwidth, scale=1]{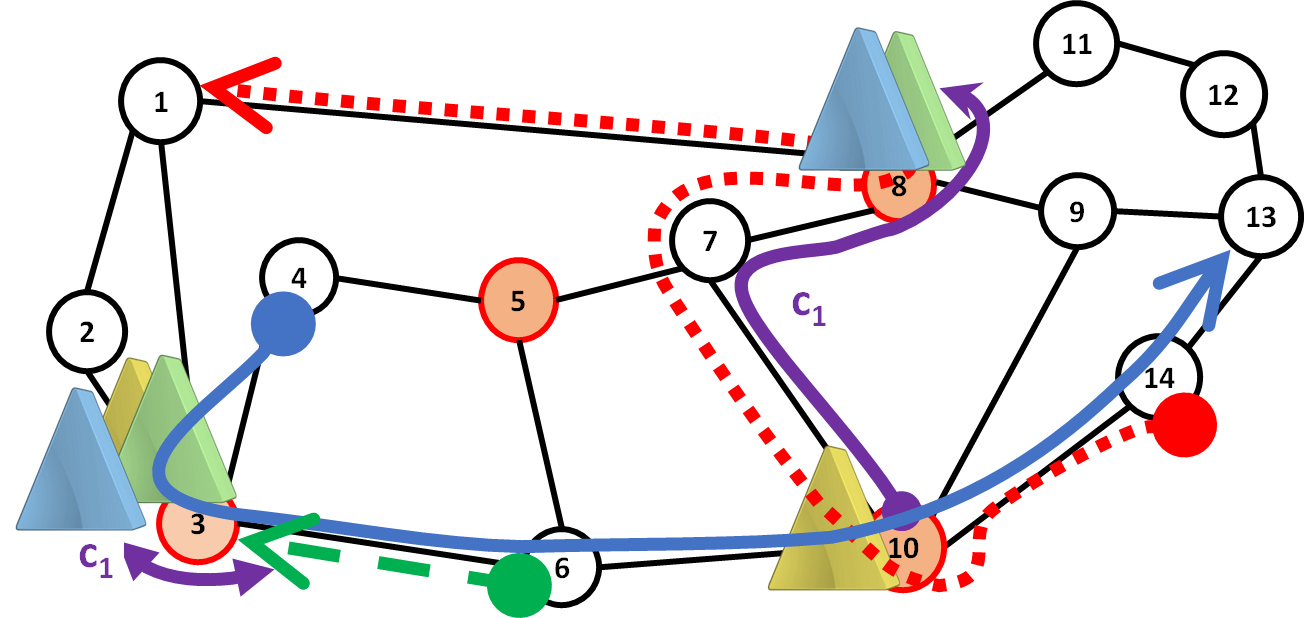}}    
  \end{tabular}
  \caption{Deploying more SC occurrence mappings reduces network resource consumption.}
  \label{configIntro}
\end{figure*}

When several network functions are configured to provide a service, we have a ``Service Chain''. The term ``service chain'' is used ``to describe the deployment of such functions, and the network operator's process of specifying an ordered list of service functions that should be applied to a deterministic set of traffic flows'' \cite{ietf_sc}. So, a ``Service Chain'' (SC) specifies a set of network functions configured in a specific order. With NFV, we can form SCs where VNFs are configured in a specific sequence that minimizes the bandwidth usage in the network (an example is discussed below).

Unfortunately, since VNFs in a single SC may need to be traversed by several distinct  traffic flows (i.e., flows requested by multiple geographically-distributed node pairs) in a specific sequence, it becomes difficult to improve network resource utilization. For example, consider Figs. \ref{configIntro}\subref{fig:a} and \ref{configIntro}\subref{fig:b}, where three traffic requests $r_1$ (from node 4 to 13), $r_2$ (from node 6 to 3), and $r_3$ (from node 14 to 1) demand SC $c_1$ composed of VNF1, VNF2, and VNF3 (to be traversed in this order VNF1$\,\to\,$VNF2$\,\to\,$VNF3).
In Fig. \ref{configIntro}\subref{fig:a}, if we consider only one mapping occurrence (or instance) for SC $c_1$, then some traffic flows (in our example, $r_3$ and $r_2$) will be ineffectively routed over long paths. Instead, as shown in Fig. \ref{configIntro}\subref{fig:b}, if we use two SC {\SCcopies} for the same SC, 
we can improve network resource utilization, at the expense of a larger number of VNFs to be deployed (or replicated) in the network to serve the same SC.
This results in a more complex problem when, in a Wide-Area Network (WAN), a large number of distributed node pairs generate traffic flows, creating heavy traffic demands. Our objective in this work is to reduce the network resource consumption for a WAN with heavy traffic demands.  


So the question is: how many SC {\SCcopies} for the same SC are required for optimal network resource utilization?

A possible (trivial) solution to the problem of SC mapping in case of multiple node pairs requiring the same SC is to use one single {\SCcopy} that would most likely lead to host SCs at a single node (e.g., a DC) which is centrally located in the network. 
However, traffic flows may have to take long paths to reach the node hosting the SC, which will result in a high network resource consumption. 

The other extreme case would be to use a distinct SC mapping per each node pair (in other words, the number of SC {\SCcopies} is equal to the number of traffic node pairs). Now, we can achieve optimal network resource utilization as each node pair will use an SC  effectively mapped along a shortest path in the network\footnote{Using the shortest path also has the added effect of reducing latency for the service chain, but this aspect is out of scope for this study.}. However,  this approach will increase the network orchestration overhead and increase capital expenditure, as there will be a large number of replicated VNF instances across nodes. To reduce excessive VNF replication, we bound the maximum number of nodes hosting VNFs.   

Intuitively, the number of SC {\SCcopies} for a good solution will be a value between these two extremes. This solution will minimize the network resource utilization while not excessively increasing the number of nodes hosting VNFs.  

A reasonable trade-off that leads to the optimal solution is difficult to calculate, as the problem of SC mapping with multiple SC {\SCcopies} results in quadratic constraints \cite{sc_related} that severely hamper the scalability of the solution. In this study, to answer the question above, we propose a two-phase solution, relying on a column-generation-based ILP model, which provides quasi-optimal solutions with reasonable computational time. Sub-optimality comes from the fact that we solve the problem in two phases: in the first phase, we group node pairs that will be forced to use the same SC {\SCcopy}; in the second phase, we run our scalable column-generation approach to find a solution starting from the grouping already performed in the first phase.   
Applying this approach over two realistic network topologies, we observe that an appropriate choice of only a small set of different SC mappings can lead to a solution very close to the  minimum theoretical bandwidth consumption, even for a full-mesh traffic demand matrix. 




The rest of this study is organized as follows. 
Section \ref{relWrk} overviews the existing literature on the SC mapping problem and remarks the novel contributions of this study. Section \ref{probDesc} formally describes the problem and its input parameters.  
Section \ref{ilp} describes the Integer Linear Program (ILP) formulation for the problem, while Section \ref{colGen} describes the quadratic column-generation-based ILP model.
Section \ref{twopm} introduces a heuristic to cluster groups of node pairs that will use the same SC {\SCcopy}; and then describes our column-generation-based ILP solution method. 
Section \ref{sim_examples} provides some illustrative examples which demonstrate  that  a limited number of SC {\SCcopies} can lead to quasi-optimal solution of the problem. Section \ref{concl} concludes the study.


\section{Related Work}
\label{relWrk}

A number of studies exist on the VNF placement and routing problem. 
Ref. \cite{sc_related} was the first to formally define the problem of VNF placement and routing. However, they developed a Quadratic Constrained Program (QCP), making it unscalable beyond small problem instances. Ref. \cite{vnf_placement_turck} studied a hybrid deployment scenario with hardware middleboxes using an ILP, but did not enforce VNF service chaining  explicitly. Ref. \cite{place_vnf_secci} used an ILP to study trade-offs between legacy and NFV-based traffic engineering but did not have explicit VNF service chaining. Ref. \cite{vnf_placement_barcellos_gaspary} modeled the problem in a DC setting using an ILP to reduce the end-to-end delays and minimize resource over-provisioning while providing a heuristic to do the same. Here too VNF service chaining is not explicitly enforced by the model. Ref. \cite{orc_vnf_boutaba} modeled the batch deployment of multiple chains using an ILP and developed heuristics to solve larger instances of the problem. However, it enforced that VNF instances of a function need to be on a single machine and restricts all chains to three VNFs. Our model does not impose such constraints, and we allow any VNF type to be placed at any node and any number of VNFs in a SC while service chaining VNFs for a SC explicitly. Ref. \cite{sc_detail} accounted for the explicit service chaining of VNFs but focused on compute resource sharing among VNFs. Ref. \cite{nicolas_vnf} used a column-generation model to solve VNF placement and routing but considered dedicated SC instances per traffic pair, hence solving the second extreme case mentioned in the introduction, which is a particular case of our approach. Ref. \cite{zhu_tnsm} also used a column-generation model to solve the dynamic VNF placement and routing problem but considered a single SC instance per SC, which as mentioned earlier will lead to a sub-optimal solution. 

Recently, there have been a few works on using multiple VNF instances for load balancing to reducing resource utilization and improve QoS. There are several differences between these and our work which we clarify below. Ref. \cite{replica_jukan} developed a load-balancing scheme for the Virtual Evolved Packet Core (vEPC) SC, given a set of pre-computed paths by replicating the instances of certain (not all) VNFs. Our approach deals with a SC in general without any prior computation of paths. Ref. \cite{replica_michstate} developed an online approach for scaling SCs by using VNF replicas and an approximate version of an offline scheme. It provided theoretical bounds for its technique; however, the method does not provide a general mapping of an SC instance to network node and VNF replicas which is done by our approach. Ref. \cite{replica_dcs_cuhk} looked at selection of NFV nodes and VNF assignment separately. Our approach does node selection and VNF assignment jointly while also holistically mapping SC instances to the allowed number of NFV nodes and VNF replicas for each SC. 

 
Our previous work \cite{ilp_report} and most existing works solve the problem for multiple SCs, but for each SC only a single {\SCcopy} of the SC is allowed. We remark again that, in the current work, we consider multiple SCs, but for each SC, multiple {\SCcopies} per SC are allowed; hence most existing works represent a particular case of our current work, where each node pair requesting an SC has its own {\SCcopy}. Further, we also consider multiple geographically-distributed node pairs which create heavily-populated (dense) traffic demands. As extending the model to multiple {\SCcopies} per SC results in quadratic constraints, we propose a novel decomposition model (column generation) for SC mapping with multiple SC {\SCcopies}, which, 
together with a traffic-grouping  heuristic, provides a scalable solution to the problem (Section \ref{twopm}). 

To the best of our knowledge, this is the first attempt to address the solution of the complete SC mapping problem (i.e., with multiple SC {\SCcopies}) over large network instances.


\section{Problem Description}
\label{probDesc}

An operator's network provides multiple services, and each service is realized by traversing a Service Chain (SC). To provide multiple services, the operator has to map corresponding SCs into network. We develop three solution approaches for this multiple SC mapping problem. First is the ILP described in Section \ref{ilp}. Second is a column-generation-based ILP (CG-ILP) detailed in Section \ref{colGen}. Finally, we solve the problem using a two-phase approach described in Section \ref{twopm}. 

\subsection{Problem Statement}
\label{probState}

Given a network topology, capacity of links, a set of network nodes with NFV support (NFV nodes), compute resources at NFV nodes, maximum number of NFV nodes that can be used, traffic flows for source-destination pairs requiring a specific SC with a  certain bandwidth demand, a set of VNFs, and a set of SCs, we determine the placement of VNFs and corresponding traffic routing to minimize network-resource (bandwidth) consumption. Note that VNFs can be shared among different SCs.

\subsection{Input Parameters}

\begin{itemize}
\item $G = (V, L)$: Physical topology of backbone network; $V$ is set of nodes and $L$ is set of links.
\item $V^{\NFV} \subseteq V$: Set of nodes that can host VNFs (NFV nodes).
\item $I_{c}$: Number of instances for SC $c$.
\item $K$: Maximum number of NFV nodes allowed to host VNFs.
\item $F$, indexed by $f$: Set of VNFs.
\item $R_{f}$: Maximum number of replicas of VNF $f$. 
\item $\ncore$: Number of CPU cores present in a NFV node.
\item $\ncoref$: Number of CPU cores per Gbps for function $f$.
\item $\ServiceChainSet$: Set of chains, indexed by $c$.
\item $n_c$: Number of VNFs in SC $\ServiceChainIdx$.
\item $\mathcal{SD}$: Set of source-destination $(v_s,v_d)$ pairs.
\item $\mathcal{SD}_c$: Set of source-destination $(v_s,v_d)$ pairs for SC $c$.
\item $D^c_{sd}$: Traffic demand between $v_s$ and $v_d$ for SC $\ServiceChainIdx$.
\item $\sigma_i(\ServiceChainIdx)$: ID of $i$th VNF in SC $\ServiceChainIdx$ where $f_{\sigma_i(\ServiceChainIdx)} \in F$.
\item $\Tfic$: Utility for translating the $i$th VNF in SC $c$ to its VNF index $f$.
\end{itemize}


To facilitate model formulation and discussion, we propose the concept of configuration ($\OrigConfigIdx$). We use the following notation for SC representation. Each SC, denoted by $\ServiceChainIdx$, is characterized by an ordered set of $n_c$ functions:
\begin{equation}
\text{[SC $\ServiceChainIdx$]} \qquad 
f_{\sigma_1(\ServiceChainIdx)} \prec f_{\sigma_2(\ServiceChainIdx)} \prec \dots \prec f_{\sigma_{n_\ServiceChainIdx}(\ServiceChainIdx)}
\end{equation}
Each deployment of SC $\ServiceChainIdx$ is defined by a set of VNF locations, a set of paths, from location of first VNF to location of last VNF, and set of traffoc flows traversing this deployment.

We generate a set of \textit{SC configurations} where each configuration ($\OrigConfigIdx$) is associated with a potential provisioning of a SC $\ServiceChainIdx$, i.e., with a potential node placement of its functions and a potential subset of traffic flows from $\mathcal{SD}_c$. 
Let $\OrigConfigSet$ be the set of configurations, and $\OrigConfigSet_\ServiceChainIdx$ be the subset of configurations associated with service chain $\ServiceChainIdx \in \ServiceChainSet$:
$\quad \OrigConfigSet = \bigcup\limits_{\ServiceChainIdx \in \ServiceChainSet} \OrigConfigSet_\ServiceChainIdx.$

Potential set of configurations for a SC $c$ is given by:
$$\OrigConfigSet_\ServiceChainIdx = \sum_{sd=1}^{N_{SD_c}} {N_{SD_c} \choose sd} \times {\{N_{V^{NFV}}\}}^{n_c} \times {P_{paths}}^{n_c-1}$$
where $sd$ is the number of number of source-destination $(v_s,v_d)$ pairs using a configuration, $N_{SD_c}$ gives the number of source-destination $(v_s,v_d)$ pairs for SC $c$, $N_{V^{NFV}}$ gives the number of NFV nodes and $P_{paths}$ refers to the number of paths from the location of $f_{\sigma_i(\ServiceChainIdx)}$ to the location of $f_{\sigma_{i+1}(\ServiceChainIdx)}$.

A chain configuration ($\OrigConfigIdx$) is characterized by the following parameters: 
\begin{itemize}
\item Traffic flows: $\asdILP = 1$ if $(v_s,v_d)$ uses configuration $\OrigConfigIdx$; 0 otherwise.
\item Location of functions: $a_{vi}^{\OrigConfigIdx} =1$ if $i$th function $f_i \in \ServiceChainIdx$ is located in $v$ in configuration $\OrigConfigIdx$; 0 otherwise.
\item Connectivity of locations: path from location of current VNF to next VNF in SC $\ServiceChainIdx$. If link $\ell$ is used in the path from location of $f_{\sigma_i(\ServiceChainIdx)}$ to location of $f_{\sigma_{i+1}(\ServiceChainIdx)}$, then $b_{i \ell}^{\OrigConfigIdx} = 1$; 0 otherwise.
\end{itemize}

\section{Integer Linear Program}
\label{ilp}

We precompute $\OrigConfigSet$, which is an input for our ILP model. ILP selects the best configuration ($\OrigConfigIdx$) based on other input parameters and constraints, and computes the route from $v_s$ (source) to first VNF of $c$ and from last VNF of $c$ to $v_d$ (destination) for each source-destination $(v_s, v_d)$ pair. 

\textbf{Variables:}
\begin{itemize}
\item $z_{\OrigConfigIdx} =1$ if configuration $\OrigConfigIdx$ is selected; 0 otherwise.
\item $\xvic =1$ if $i$th function of $\ServiceChainIdx$ is located in $v$; 0 otherwise.
\item $y^{f_1(\ServiceChainIdx), sd}_{\ell} =1$ if $\ell$ is on path from $v_s$ to location of first VNF in $\ServiceChainIdx$; 0 otherwise.
\item $y^{f_{n_c}(\ServiceChainIdx), sd}_{\ell} =1$ if $\ell$ is on path from location of last VNF in $\ServiceChainIdx$ to $v_d$; 0 otherwise.
\item $h_v =1$ if $v$ is used as a location for a VNF; 0 otherwise.
\end{itemize}

\noindent
\textbf{Objective:} Minimize bandwidth consumed:
\begin{multline} \min \quad 
     \sum\limits_{\ServiceChainIdx \in \ServiceChainSet} \> \sum\limits_{{\OrigConfigIdx} \in \OrigConfigSet_\ServiceChainIdx} \>    
     \overbrace{ \left(   \sum\limits_{(s,d) \in \mathcal{SD}}   \> D^c_{sd}  \right) }^{\text{Overall traffic using } \ServiceChainIdx} 
     \overbrace{ \left( \sum\limits_{\ell \in L} \> \sum\limits_{i \in I} \asdILP b_{i\ell}^{\OrigConfigIdx} \right)}^{ \substack{\text{Number of links} \\  \text{ in the route of } \ServiceChainIdx}}\zILP  + \\
     \sum\limits_{\ServiceChainIdx \in \ServiceChainSet} \> 
     \sum\limits_{\ell \in L} \> \sum\limits_{(s,d) \in \mathcal{SD}}   \> D^c_{sd}  \left( y^{f_1(\ServiceChainIdx), sd}_{\ell} + y^{f_{n_c}(\ServiceChainIdx), sd}_{\ell} \right)    
\end{multline}

Total bandwidth consumed in placing multiple SCs depends on configurations ($\OrigConfigIdx$'s)  selected for each SC  $\ServiceChainIdx$. Each $\OrigConfigIdx$ for $\ServiceChainIdx$ locates VNFs of $\ServiceChainIdx$ and gives the route to traverse these VNF locations. So, bandwidth consumed when going from $v_s$ to $v_d$ and traversing the SC depends on selected $\OrigConfigIdx$. 

\textbf{Constraints}:
\begin{alignat}{2}
 & \sum\limits_{{\OrigConfigIdx} \in \OrigConfigSet_\ServiceChainIdx} {\zILP} \leq \ncopyc  
 && \hspace*{-3.5cm} \ServiceChainIdx \in \ServiceChainSet \label{eq1:at_most_ncopyc_configs_for_SFCc} \\
& \sum\limits_{\ServiceChainIdx \in \ServiceChainSet} \> \sum\limits_{{\OrigConfigIdx} \in \OrigConfigSet_\ServiceChainIdx}  \> \sum\limits_{i=1}^{n_{\SFCidx}} \> 
 \Tfic a_{v i}^{\OrigConfigIdx}  \>  \zILP \leq M x_{vf}
&& \hspace*{-1.5cm} f \in F, v \in V^{\NFVI} \label{eq1:is_f_in_v} \\
& \sum\limits_{\ServiceChainIdx \in \ServiceChainSet} \> \sum\limits_{{\OrigConfigIdx} \in \OrigConfigSet_\ServiceChainIdx}  \> \sum\limits_{i=1}^{n_{\SFCidx}} \> 
 \Tfic a_{v i}^{\OrigConfigIdx}  \>  \zILP \geq x_{vf}
&& \hspace*{-1.5cm} f \in F, v \in V^{\NFVI} \label{eq1:is_f_in_v2} \\
& \sum\limits_{v \in V^{\NFVI}} x_{vf} \leq \nlocationf 
&& \hspace*{-3.5cm} f \in F \label{eq1:count_locationf} \\\
& M h_v \geq \sum\limits_{f \in F} x_{vf} \geq h_v && \hspace*{-3.cm} v \in V^{\NFV} \label{eq1:vnf_location} \\
& \sum\limits_{v \in V^{\NFV}} h_v \leq K \label{eq1:k_nfv_nodes}\\
& \sum\limits_{\ServiceChainIdx \in \ServiceChainSet} \>   \sum\limits_{\OrigConfigIdx \in  \OrigConfigSet_\ServiceChainIdx} \>
     \> \sum\limits_{(v_s, v_d) \in \mathcal{SD}}  D_{sd}^c \> {\asdILP}* \> 
&& \nonumber \\     
 &    \left( \sum\limits_{f \in F} \> \sum\limits_{i=1}^{n_c} \Tfic \lambdaf  a_{v i}^{\OrigConfigIdx} \right) \> \zILP \leq \Ncore
&& \hspace*{-1cm} v \in V^{\NFV}  \label{eq1:capa_cores} \\
&   \sum\limits_{\ServiceChainIdx \in \ServiceChainSet} \>  \sum\limits_{(v_s, v_d) \in \mathcal{SD}}  \> D^{\ServiceChainIdx}_{sd}* \> 
&& \nonumber \\
&  \left(    y^{f_1(\ServiceChainIdx), sd}_{\ell} 
        + y^{f_{n_c}(\ServiceChainIdx), sd}_{\ell}     
        + \sum\limits_{{\OrigConfigIdx} \in \OrigConfigSet_{\ServiceChainIdx}} \> {\asdILP} \> \zILP \sum\limits_{i =1}^{n_{\ServiceChainIdx} - 1} b_{i \ell}^{\OrigConfigIdx} \>    
    \right)   && \nonumber \\
& \qquad \qquad \qquad \qquad 
  \leq \textsc{cap}_{\ell}  
&& \hspace*{-2cm}  \ell \in L \label{eq1:capacity} \\
& \sum\limits_{{\OrigConfigIdx} \in \OrigConfigSet_\ServiceChainIdx} \asdILP z_{\OrigConfigIdx} = 1  
&& \hspace*{-4cm} \ServiceChainIdx \in \ServiceChainSet, (v_s,v_d) \in \mathcal{SD}: D_{sd}^{\ServiceChainIdx} > 0 \label{eq1:one_path_per_sd_c}
\end{alignat}

Constraints \eqref{eq1:at_most_ncopyc_configs_for_SFCc} guarantee that we select exactly $\ncopyc$ configurations for SC  $\ServiceChainIdx$  
and force $\ServiceChainIdx$ to have $\ncopyc$ {\SCcopies}. 
Each $\OrigConfigIdx$ is associated with a set of $a^{\OrigConfigIdx}_{vi}$ required to be consistent with $x_{vf}$, which is resolved by Eqs. \eqref{eq1:is_f_in_v}, \eqref{eq1:is_f_in_v2} where $\Tfic$ is to find the VNF $f$ at sequence $i$ in SC $c$. Eq. \eqref{eq1:count_locationf} is used to limit the number of VNF replicas. Eq. \eqref{eq1:vnf_location} is used to keep track of NFV nodes used for hosting VNFs while Eq. \eqref{eq1:k_nfv_nodes} limits the number of NFV nodes allowed to host VNFs.
Constraints \eqref{eq1:capa_cores} ensure that each NFV node has a sufficient number of CPU cores for hosting $f$. Eq. \eqref{eq1:capacity} constrains link capacity. Eq. \eqref{eq1:one_path_per_sd_c} enforces that, for each source-destination pair $(v_s,v_d)$ requesting SC $c$, there is exactly one configuration $\OrigConfigIdx$.

\begin{alignat}{2}     
& \text{\textbf{Route from} } v_s \text{ \textbf{to first function location:}} \nonumber \\
& \sum\limits_{\OrigConfigIdx \in \OrigConfigSet_{\ServiceChainIdx}} {\asdILP} \> a_{v_s, 1}^{\OrigConfigIdx} \zILP +  \sum\limits_{\ell \in \omega^+{(v_s)}} y^{f_1(\ServiceChainIdx), sd}_{\ell}  = 1
&& \nonumber \\
& &&  \hspace*{-5.cm}  \ServiceChainIdx \in \ServiceChainSet, (v_s,v_d) \in \mathcal{SD}: D_{sd}^{\ServiceChainIdx} > 0 \label{eq1:link_from_source_to_ingress} \\
&  \sum\limits_{\OrigConfigIdx \in \OrigConfigSet_{\ServiceChainIdx}} {\asdILP} \>  a_{v 1}^{\OrigConfigIdx} \zILP - \sum\limits_{\ell \in \omega^-{(v)}} y^{f_1(\ServiceChainIdx), sd}_{\ell}  \leq 0
&&  \nonumber \\
& && \hspace*{-5.cm}  \ServiceChainIdx \in \ServiceChainSet, (v_s,v_d) \in \mathcal{SD}: D_{sd}^{\ServiceChainIdx} > 0, \nonumber \\
& && \hspace*{-5.cm} v \in V^{\NFVI}  \setminus \{ v_s \} \label{eq1:to_ensure_NFV_1_placement} \\
&  \sum\limits_{\OrigConfigIdx \in \OrigConfigSet_{\ServiceChainIdx}} {\asdILP} \>  a_{v 1}^{\OrigConfigIdx} \zILP +
        \sum\limits_{\ell \in \omega^+{(v)}} y^{f_1(\ServiceChainIdx), sd}_{\ell}
     -  \sum\limits_{\ell \in \omega^-{(v)}} y^{f_1(\ServiceChainIdx), sd}_{\ell} 
     = 0
&& \nonumber \\
& && \hspace*{-5.cm}  \ServiceChainIdx \in \ServiceChainSet, (v_s,v_d) \in \mathcal{SD}: D_{sd}^{\ServiceChainIdx} > 0, \nonumber \\
& && \hspace*{-5.cm} v \in V^{\NFVI}  \setminus \{ v_s \}        \label{eq1:places_NFV} \\
&     \sum\limits_{\ell \in \omega^+{(v)}} y^{f_1(\ServiceChainIdx), sd}_{\ell}
     -  \sum\limits_{\ell \in \omega^-{(v)}} y^{f_1(\ServiceChainIdx), sd}_{\ell}
     = 0
      && \nonumber \\
      & && \hspace*{-5.cm}  \ServiceChainIdx \in \ServiceChainSet, (v_s,v_d) \in \mathcal{SD}: D_{sd}^{\ServiceChainIdx} > 0, \nonumber \\
      & && \hspace*{-5.cm} v \in V \setminus (V^{\NFVI}  \cup \{ v_s \})         \label{eq1:places_non_NFV} 
\end{alignat} 

We assume that an unique route exists from $v_s$ to first VNF location. This is imposed by selecting exactly one outgoing link from $v_s$ unless first VNF is located at 
$v_s$. We account for these scenarios using Eq. \eqref{eq1:link_from_source_to_ingress}. To find the route from $v_s$ to first VNF, flow conservation needs to be enforced at the intermediate nodes which may or may not have NFV support. Eqs. \eqref{eq1:places_NFV} and \eqref{eq1:places_non_NFV} enforce flow-conservation constraints at nodes with and without NFV support, respectively.

\begin{alignat}{2}
& \text{\textbf{Route from last function location to }} v_d \text{\textbf{:}}\nonumber \\
& \sum\limits_{\OrigConfigIdx \in \OrigConfigSet_{\ServiceChainIdx}} {\asdILP} \> a_{v_d, n_c}^{\ConfigIdx} \zILP +  \sum\limits_{\ell \in \omega^-{(v_d)}} y^{f_{n_c}(\ServiceChainIdx), sd}_{\ell} 
    = 1
&&  \nonumber \\
& && \hspace*{-6.cm}  \ServiceChainIdx \in \ServiceChainSet, (v_s,v_d) \in \mathcal{SD}: D_{sd}^{\ServiceChainIdx} > 0  \label{eq1:link_from_egress_to_destination} \\      
& \sum\limits_{\OrigConfigIdx \in \OrigConfigSet_{\ServiceChainIdx}} {\asdILP} \> a_{v, n_c}^{\OrigConfigIdx} \zILP - \sum\limits_{\ell \in \omega^+{(v)}} y^{f_{n_c}(\ServiceChainIdx), sd}_{\ell} \leq 0  
&&   \nonumber \\
& && \hspace*{-6.cm} \ServiceChainIdx \in \ServiceChainSet,  (v_s, v_d) \in \mathcal{SD}: D_{sd}^{\ServiceChainIdx} > 0, \nonumber \\
& && \hspace*{-6.cm} v \in V^{\NFVI}  \setminus \{ v_d \}  
      \label{eq1:to_ensure_NFV_N_placement} \\ 
&   \sum\limits_{\OrigConfigIdx \in \OrigConfigSet_{\ServiceChainIdx}} {\asdILP} \> a_{v, n_c}^{\OrigConfigIdx} \zILP -   \sum\limits_{\ell \in \omega^+{(v)}} y^{f_{n_c}(\ServiceChainIdx), sd}_{\ell} 
    +  \sum\limits_{\ell \in \omega^-{(v)}} y^{f_{n_c}(\ServiceChainIdx), sd}_{\ell}  = 0 && \nonumber \\
& && \hspace*{-6.cm}  \ServiceChainIdx \in \ServiceChainSet, (v_s, v_d) \in \mathcal{SD}: D_{sd}^{\ServiceChainIdx} > 0,  \nonumber \\
& && \hspace*{-6.cm} v \in V^{\NFVI}  \setminus \{ v_d \} \label{eq1:places_NFV_destination} \\
&      \sum\limits_{\ell \in \omega^+{(v)}} y^{f_{n_c}(\ServiceChainIdx), sd}_{\ell} 
     -   \sum\limits_{\ell \in \omega^-{(v)}} y^{f_{n_c}(\ServiceChainIdx), sd}_{\ell}  = 0 && \nonumber \\
    & && \hspace*{-6.cm}  \ServiceChainIdx \in \ServiceChainSet, (v_s, v_d) \in \mathcal{SD}: D_{sd}^{\ServiceChainIdx} > 0,  \nonumber \\
    & && \hspace*{-6.cm} v \in V \setminus (V^{\NFVI}  \cup \{ v_d \}) \label{eq1:places_not_NFV_destination}    
\end{alignat}

Eq. \eqref{eq1:link_from_egress_to_destination} selects one incoming link to $v_d$ to ensure a route to $v_d$. For cases where last VNF is placed at destination node, we use Eq. \eqref{eq1:to_ensure_NFV_N_placement}. Eqs. \eqref{eq1:places_NFV_destination} and \eqref{eq1:places_not_NFV_destination} enforce flow conservation at nodes with and without NFV support, respectively.

\section{Column Generation - ILP}
\label{colGen}

Pre-computing all configurations becomes computationally expensive for large networks. As the number of configurations grows with network size, the problem fits naturally in the column-generation framework \cite{jaumard_colgen_rwa}. 

Column generation (\textbf{CG}) is a decomposition technique, where the problem (called Master Problem-\textbf{MP}) to be solved is divided into two sub-problems: restricted master problem (\textbf{RMP}) (selection of the best configurations) and pricing problems
 (\textbf{PP\_SC($ \ServiceChainIdx$)})$_{\ServiceChainIdx \in \ServiceChainSet}$ (configuration generators for each chain). 
CG process involves solving \textbf{RMP}, querying the dual values of \textbf{RMP} constraints, and using them for \textbf{PP\_SC($ \ServiceChainIdx$)} objective. Each improving solution (i.e., with a negative reduced cost) of \textbf{PP\_SC($ \ServiceChainIdx$)} is added to \textbf{RMP}, and previous step is repeated until optimality condition is reached (\cite{jaumard_colgen_rwa,jau17RWA}), with \textbf{PP\_SC($ \ServiceChainIdx$)} explored in a round-robin fashion.

The advantage here is that we do not have to precompute configurations. \text{CG} generates a column (here, a configuration) by itself, adds them to \textbf{RMP}  and solves \textbf{RMP}. This set of steps is repeated until reduced cost becomes non-negative ($\redcost \geq 0$). We convert the final \textbf{RMP} to an ILP and solve to get integer solution.
\textbf{RMP} selects the best $\ConfigIdx \in \ConfigSet_\ServiceChainIdx$ for each SC $\ServiceChainIdx$. Also it finds a route from $v_s$ (source) to first VNF of $\ServiceChainIdx$ and from last VNF of $\ServiceChainIdx$ to $v_d$ (destination).

\begin{figure}[htb]
\begin{center}
  \subfloat[][A first configuration ($\ConfigIdx_1$) for $c$]{\label{fig:Config1}\includegraphics[scale=.25]{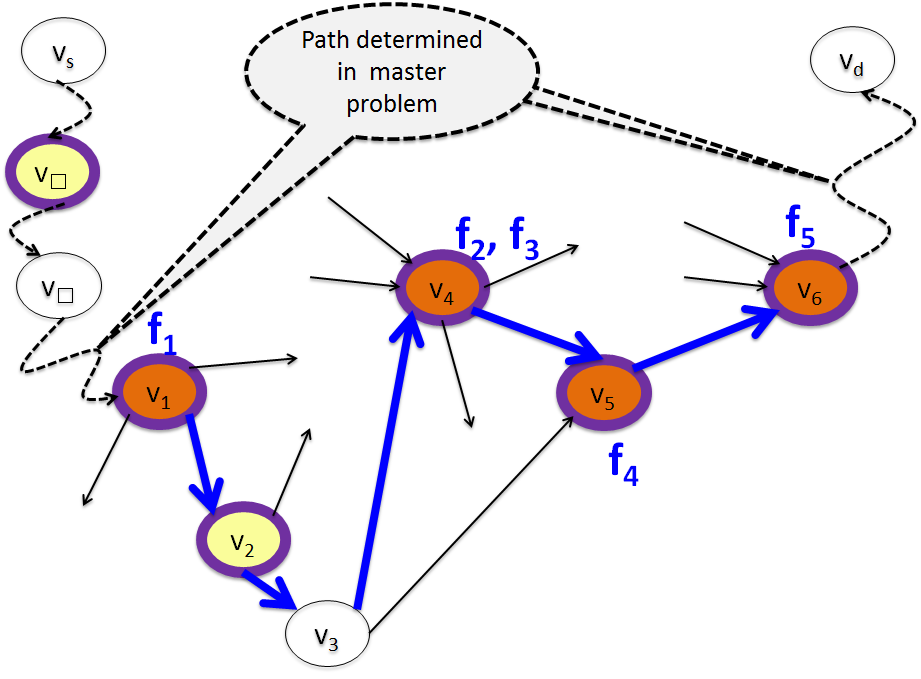}} \\
  \subfloat[][A second configuration ($\ConfigIdx_2$) for $c$]{\label{fig:Config2}\includegraphics[scale=.25]{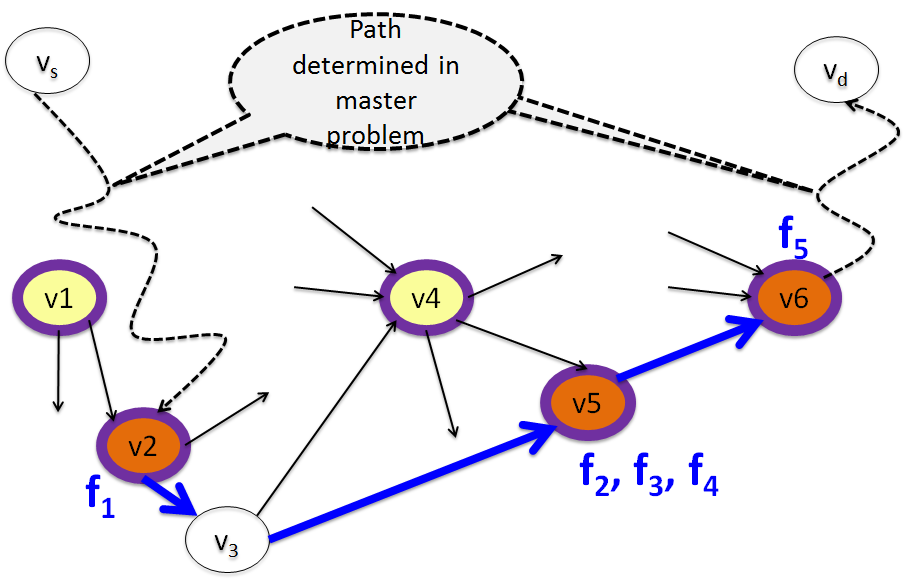}}
\end{center}
\caption{Two configuration examples for  chain $c = (f_1 \prec f_2 \prec f_3 \prec f_4 \prec f_5)$.}
\label{fig:service_chain}	
\end{figure}

An illustration of the constraint splitting between \textbf{RMP} and \textbf{PP\_SC($ \ServiceChainIdx$)} is depicted in Fig. \ref{fig:service_chain}. Nodes circled in purple are NFV nodes, yellow nodes do not host VNFs at present but have NFV support, and orange nodes currently host VNFs. Figure \ref{fig:service_chain}\subref{fig:Config1} has $f_1$ located at $v_1$. When a different configuration is selected in Fig. \ref{fig:service_chain}\subref{fig:Config2} and $f_1$ is located at $v_2$, then \textbf{RMP} finds the path from $v_s$ to location of $f_1$. Similarly, \textbf{RMP} finds the path from last VNF to $v_d$, i.e., $f_5$ to $v_d$ here. 

\subsection{Reduced Master Problem (RMP)}
\label{colGenRMP}

\textbf{Objective:} Minimize bandwidth consumed:
\begin{multline} \min \quad 
     \sum\limits_{\ServiceChainIdx \in \ServiceChainSet} \> \sum\limits_{{\OrigConfigIdx} \in \OrigConfigSet_\ServiceChainIdx} \>
     \underbrace{ 
   		 \left(   \sum\limits_{(s,d) \in \mathcal{SD}}   \> D^c_{sd}  \right) 
         \left( \sum\limits_{\ell \in L} \> \sum\limits_{i \in I} \underbrace{ \asdILP b_{i\ell}^{\OrigConfigIdx}}_{y_{i \ell}^{\OrigConfigIdx, sd}} \right)
                   }_{\textsc{cost}_{\OrigConfigIdx}} 
                   \zILP  + \\
     \sum\limits_{\ServiceChainIdx \in \ServiceChainSet} \> 
     \sum\limits_{\ell \in L} \> \sum\limits_{(s,d) \in \mathcal{SD}}   \> D^c_{sd}  \left( y^{f_1(\ServiceChainIdx), sd}_{\ell} + y^{f_{n_c}(\ServiceChainIdx), sd}_{\ell} \right)    
\end{multline}

The formulation for the reduced master problem is same as the ILP in Section {\ref{ilp}}. However, the \textbf{RMP} is solved as a Linear Program (LP) for the duration of the CG. After the CG solves the RMP optimally (when $\redcost \geq 0$), we solve the final RMP as an ILP to get integer solution.

\subsection{Pricing Problem: PP$(\SFCidx)$}
\label{colGenPP}

\textbf{PP\_SC($ \ServiceChainIdx$)} generates configurations. Here, we discover that the configuration ($\OrigConfigIdx$) structure results in \textbf{quadratic} constraints since we have to also determine which traffic flows will make up the configuration. \textbf{Quadratic} expressions can be seen in Eqs. \eqref{eq2:red_cost}, \eqref{eq2:PP_node_capacity}, and \eqref{eq2:PP_link_capacity}. We \textbf{linearize} these constraints using Eqs. \eqref{eq2:and1} and \eqref{eq2:and2}, however, the performance of the CG is still affected. 


\noindent
\textbf{Objective:} Minimize reduced cost of variable $z_{\OrigConfigIdx}$ (after linearization): 
\vspace{-.2cm}
\begin{multline}
\label{eq2:red_cost}
\text{[\textbf{PP\_SC($\ServiceChainIdx$)}]} \qquad 
 \redcost_{\OrigConfigIdx} = \cost_{\OrigConfigIdx}
+ u^{\eqref{eq1:at_most_ncopyc_configs_for_SFCc}} \\
 + \sum\limits_{v \in V^{\NFVI}} \> \sum\limits_{f \in F} \sum\limits_{i=1}^{n_{\SFCidx}} \> u^{\eqref{eq1:is_f_in_v}}_{fv} \Tfic a_{v i} 
 - \sum\limits_{v \in V^{\NFVI}} \> \sum\limits_{f \in F} \sum\limits_{i=1}^{n_{\SFCidx}} \> u^{\eqref{eq1:is_f_in_v2}}_{fv} \Tfic a_{v i} \\
+ \sum\limits_{v \in V^{\NFV}} \> u^{\eqref{eq1:capa_cores}}_{v}  \>{\sum\limits_{(v_s, v_d) \in \SD} } D_{sd}^{\ServiceChainIdx} \> 
   \sum\limits_{i =1}^{n_c}   \Tfic \lambdaf \csdvi \\
+ \sum\limits_{\ell \in L} \sum\limits_{(v_s, v_d) \in \SD} u^{\eqref{eq1:capacity}}_{\ell} D_{sd}^c 
                  \sum\limits_{i =1}^{n_{\ServiceChainIdx} - 1} \qsdli 
   -  \sum\limits_{(v_s, v_d) \in \SD} \> u_{sd}^{\eqref{eq1:one_path_per_sd_c}} {\asdPP}\\
                  - \sum\limits_{(v_s, v_d) \in \SD} u_{sd}^{\eqref{eq1:link_from_source_to_ingress}} \csdvsrcone 
                  + \sum\limits_{(v_s, v_d) \in \SD} \sum\limits_{v \in V^{\NFVI} \setminus \{v_s \}} u_{sd,v}^{\eqref{eq1:to_ensure_NFV_1_placement}} \csdvone  \\
                   - \sum\limits_{(v_s, v_d) \in \SD} \sum\limits_{v \in V^{\NFVI} \setminus \{v_s \}} u_{sd,v}^{\eqref{eq1:places_NFV}} \csdvone \\
                  - \sum\limits_{(v_s, v_d) \in \SD} u_{sd}^{\eqref{eq1:link_from_egress_to_destination}} \csdvdestlast 
                  + \sum\limits_{(v_s, v_d) \in \SD} \sum\limits_{v \in V^{\NFVI} \setminus \{v_s \}} u_{sd,v}^{\eqref{eq1:to_ensure_NFV_N_placement}} \csdvlast \\
                   - \sum\limits_{(v_s, v_d) \in \SD} \sum\limits_{v \in V^{\NFVI} \setminus \{v_s \}} u_{sd,v}^{\eqref{eq1:places_NFV_destination}} \csdvlast              
\end{multline}
where $u^{\eqref{eq1:at_most_ncopyc_configs_for_SFCc}}$, $u_{fv}^{\eqref{eq1:is_f_in_v}}$, $u_{fv}^{\eqref{eq1:is_f_in_v2}}$, $u^{\eqref{eq1:capa_cores}}_{v}$, $u^{\eqref{eq1:capacity}}_{\ell}$, $u^{\eqref{eq1:one_path_per_sd_c}}_{\ell}$, $u_{sd}^{\eqref{eq1:link_from_source_to_ingress}}$, $u_{sd,v}^{\eqref{eq1:to_ensure_NFV_1_placement}}$, $u_{sd,v}^{\eqref{eq1:places_NFV}}$, $u_{sd}^{\eqref{eq1:link_from_egress_to_destination}}$, 
     $u_{sd,v}^{\eqref{eq1:to_ensure_NFV_N_placement}}$ and $u_{sd,v}^{\eqref{eq1:places_NFV_destination}}$ are dual variables associated with Eqs. \eqref{eq1:at_most_ncopyc_configs_for_SFCc}, \eqref{eq1:is_f_in_v}, \eqref{eq1:is_f_in_v2}, \eqref{eq1:capa_cores}, \eqref{eq1:capacity}, \eqref{eq1:one_path_per_sd_c},    
\eqref{eq1:link_from_source_to_ingress}, \eqref{eq1:to_ensure_NFV_1_placement}, \eqref{eq1:places_NFV}, \eqref{eq1:link_from_egress_to_destination}, \eqref{eq1:to_ensure_NFV_N_placement} and \eqref{eq1:places_NFV_destination} respectively.  

\noindent 
\textbf{Variables:}
\begin{itemize}
\item ${\asdPP} =1$ if configuration $\OrigConfigIdx$ to be generated contains node pair $(v_s, v_d)$ requiring $\SFCidx$; 0 otherwise.
\item $\csdvi = \asdPP \> \avi = 1$ if node pair $(v_s, v_d)$ is provisioned using the provisioning of $\SFCidx$ / placement function  ($i$th function of SFC $\SFCidx$ in location $v$) of the configuration under construction; 0 otherwise 
\item $\qsdli = \asdPP \> \bil = 1$ if node pair $(v_s, v_d)$ is provisioned using the provisioning of $\SFCidx$ (with link $\ell$ being used in the path from the location of the $i$th function to the location of the $(i+1)$th function) / placement function  of the configuration under construction; 0 otherwise
\end{itemize}

\noindent 
\textbf{Constraints:}
\begin{alignat}{2}
& \sum\limits_{(v_s, v_d) \in \SD} \> D^{\ServiceChainIdx}_{sd} \> \sum\limits_{i=1}^{n_c}  
           \lambdaf \Tfic \underbrace{ {\asdPP} \> a_{vi}}_{ {\csdvi} } \leq \Ncore && \nonumber\\
& &&  \hspace*{-2.cm}  v \in V^{\NFV} \label{eq2:PP_node_capacity} \\
& \sum\limits_{(v_s,v_d) \in \SD} D^{c}_{sd} \sum\limits_{i=1}^{n_c - 1} \underbrace{ \asdPP b_{i \ell} }_{ {\qsdli }} \leq \CAP_{\ell}    \quad 
&&  \hspace*{-1.cm}    \ell \in L \label{eq2:PP_link_capacity}
\end{alignat}

Eq. \eqref{eq2:PP_node_capacity} enforces a capacity constraints in CPU cores on all NFV nodes while Eq. \eqref{eq2:PP_link_capacity} imposes link capacity.
\vspace{-.15cm}
\begin{alignat}{2}
& \sum\limits_{v \in V^{\NFV}} a_{v i} = 1
&& \myspace \hspace*{-1.cm}  i = 1, 2, \dots, n_{\ServiceChainIdx}    \label{eq2:each_function} \\
& \csdvi = a_{vi} \land \asdPP && \hspace*{-2.cm}(v_s, v_d) \in \mathcal{SD}: D_{sd}^{\ServiceChainIdx} > 0, \nonumber \\
& && \hspace*{-2.cm} v \in V^{\NFVI}, i = 1, 2, \dots, n_{\ServiceChainIdx} \label{eq2:and1}\\
& \qsdli = b_{i \ell} \land \asdPP && \hspace*{-2.5cm} (v_s, v_d) \in \mathcal{SD}: D_{sd}^{\ServiceChainIdx} > 0, \nonumber \\
& && \hspace*{-2.5cm} \ell \in L, i = 1, 2, \dots, n_{\ServiceChainIdx} - 1 \label{eq2:and2}\\
&     \sum\limits_{\ell \in \omega^-(v)}  b_{1, \ell} \leq 1 - a_{v,1}  
&& \myspace  v \in V^{\NFVI}   \label{eq2:incoming1} \\
&     \sum\limits_{\ell \in \omega^+(v)}  b_{n_c -1, \ell} \leq 1 - a_{v,n_c}  
&& \myspace v \in V^{\NFVI}  \label{eq2:outgoinglast} \\
&    \sum\limits_{\ell \in \omega^+(v)} b_{i \ell} 
    - \sum\limits_{\ell \in \omega^-(v)}  b_{i \ell}
      = a_{v i} - a_{v, i+1} \nonumber \\
& && \hspace*{-3.cm} v \in V^{\NFVI}, i = 1, 2, \dots, n_{\ServiceChainIdx} - 1 \label{eq2:flowNFV} \\
& \sum\limits_{\ell \in \omega^+(v)} b_{i \ell} - \sum\limits_{\ell \in \omega^-(v)} b_{i \ell} = 0 && \nonumber \\
& && \hspace*{-3.cm} v \in V \setminus V^{\NFVI},  i = 1, 2, \dots, n_{\ServiceChainIdx} - 1 \label{eq2:flow} 
\end{alignat} 

Eq. \eqref{eq2:each_function} ensures that each VNF in SC  $\ServiceChainIdx$  is placed exactly once. 
Eqs. \eqref{eq2:and1}\footnote{
Linearization details:
$$ 
\forall (v_s, v_d) \in \mathcal{SD}: D_{sd}^{\ServiceChainIdx} > 0, 
\quad
\forall v \in V^{\NFVI}, 
\\ \quad
i = 1, 2, \dots, n_{\ServiceChainIdx} 
$$

Eq. \eqref{eq2:and1} can be linearly represented as below. 
\begin{equation}
\begin{split}
\csdvi \leq a_{vi}
\end{split}
\end{equation}
\begin{equation}
\begin{split}
\csdvi \leq \asdPP 
\end{split}
\end{equation}
\begin{equation}
\begin{split}
a_{vi} + \asdPP - 1 \leq \csdvi
\end{split}
\end{equation}
Eq. \eqref{eq2:and2} can also be similarly represented.
}  and \eqref{eq2:and2} introduce the variables to linearize the model. Eq. \eqref{eq2:incoming1} ensures that, if $f_1(\ServiceChainIdx)$ is located in $v$, there is no flow $b$ that is incoming to $v$. Eqs. \eqref{eq2:flowNFV} and \eqref{eq2:flow} are flow-conservation constraints: Eq. \eqref{eq2:flowNFV} for nodes with NFV support and Eq. \eqref{eq2:flow} for other nodes. Eq. \eqref{eq2:outgoinglast} ensures that, if $f_{n_c}(\ServiceChainIdx)$ is located in $v$, there is no flow $b$ that is outgoing $v$.

\vspace{-.5cm}
\subsection{Solution Scheme}
\label{solScheme}

The \textbf{PP\_SC($\ServiceChainIdx$)} are solved in a round-robin fashion, and the final \textbf{RMP} is solved as an ILP, as in \cite{jaumard_colgen_rwa,jau17RWA}.

\section{Two-Phase Model}
\label{twopm} 

As already mentioned,  we are solving this problem considering that each SC request chooses to map to one of multiple {\SCcopies}, which leads the model discussed in Section \ref{colGen} to have quadratic constraints, reducing the scalability of the model. So, to avoid quadratic constraints, we develop a new solution approach consisting of two phases:
\begin{itemize}
    \item Phase 1: We fix the number $N_c$ of {\SCcopies} accepted per SC ($N_c$ can go from 1 up to the number of demands for that SC), and then we group the traffic requests in $N_c$ groups of requests. All the requests in a group are forced to use the same SC {\SCcopy} (Section \ref{sptg}). Then we pass the $N_c$ {\SCcopies} as distinct SCs to the next phase.
    \item Phase 2: We solve the SC mapping problem with one single {\SCcopy} per SC based on the inputs of Phase 1. The solution of this simplified (linear, yet still very complex) problem  is based on a column-generation-based decomposition model (Section \ref{phase2cg}). 
\end{itemize}

As a result of Phase 1, we no longer have to account for traffic flows as part of a configuration. This happens because we partition the traffic flows in Phase 1, and so it becomes much easier to find the best possible configuration for each partition in the second phase. For the two-phase-model, the configuration is $\ConfigIdx$.  
A chain configuration $\ConfigIdx$ in the two-phase model is characterized by the following parameters: 
\begin{itemize}
\item Location of functions: $a_{vi}^{\ConfigIdx} =1$ if $i$th function $f_i \in \ServiceChainIdx$ is located in $v$ in configuration; 0 otherwise.
\item Connectivity of locations: path from location of current VNF to next VNF in SC $\ServiceChainIdx$. If link $\ell$ is used in the path from location of $f_{\sigma_i(\ServiceChainIdx)}$ to location of $f_{\sigma_{i+1}(\ServiceChainIdx)}$, then $b_{i \ell}^{\ConfigIdx} = 1$; 0 otherwise.
\end{itemize}

\vspace{-.2cm}
\subsection{Phase 1: Shortest-Path Traffic Grouping (SPTG) Heuristic }
\label{sptg} 

Now, we propose a Shortest-Path Traffic Grouping (SPTG) heuristic, which forms $N_c$ groups of node pairs for each SC (given by $SD_c$), to be given as input to  the decomposition model in Section \ref{colGen} that will treat them as distinct SC and  decide the best SC mapping for each of the $N_c$ node-pair groups. As a result, we will have a solution mapping multiple SC {\SCcopies} per SC.  

The logic of the SPTG algorithm is that groups are formed among node pairs that share links along their shortest path(s). SPTG is designed to make the largest flows take shortest paths, the intuition being that, if largest flows take shortest paths, network resource consumption will be reduced. Details of SPTG approach can be found in Algorithm \ref{algo:sptg}.

\def \clusterG {\textsc{group}}
\def \groupSD {\textsc{cluster}_{sd}}

If Algorithm \ref{algo:sptg} terminates with $SD_c^{\textsc{left}}=\emptyset$ and a number of groups that is $< N_c$, partition some of the groups in order to reach $N_c$ groups.



\begin{algorithm}
\caption{SPTG$(c)$}\label{algo:sptg}
\begin{algorithmic}[1] 
\Require $G$, $SD_c$, $N_c$
\Ensure \textsc{partition} $\leftarrow$ partition of node pairs $(v_s,v_d)$ into groups
\State \textsc{partition} $\leftarrow \emptyset$ 
\State $numberOfGroups \leftarrow 0$
\State $SD_{c}^{\textsc{left}} \leftarrow SD_c$ \Comment{list of $(v_s,v_d)$ for $c$}
\State $bigFlow \leftarrow largestFlow(SD_{c}^{\textsc{left}})$ \Comment{selects largest flow in $SD_{c}^{\textsc{left}}$ }
\While{$numberOf Groups < N_c \And SD_{c}^{\textsc{left}} \neq \emptyset$}
\For {$(v_s, v_d)$ in $G$}
\State $\groupSD \leftarrow$ set of traffic pairs whose shortest path passes through $(v_s,v_d)$
\EndFor
\State $largestCluster \leftarrow \max\limits_{(v_s, v_d): D_{sd}^c > 0} \groupSD \And bigFlow \in \groupSD$ 
\State $SD_{c}^{\textsc{left}} \leftarrow SD_{c}^{\textsc{left}} \setminus largestCluster$  \Comment{remove traffic pairs of $largestCluster$ from $SD_{c}^{\textsc{left}}$}
\State Add $largestCluster$ to \textsc{partition} 
\State $numberOfGroups \leftarrow numberOfGroups + 1$
\State $bigFlow \leftarrow largestFLow(SD_{c}^{\textsc{left}})$
\EndWhile
\If{$SD_{c}^{\textsc{left}} \neq \emptyset$}
\For{$trafficPair \in SD_{c}^{\textsc{left}}$}
\State add $trafficPair$ to  $\clusterG \in$ \textsc{partition}, such that the $(v_s,v_d)$ associated with $\clusterG$ provides the shortest path for provisioning $trafficPair$
\EndFor
\EndIf
\end{algorithmic}
\end{algorithm}

\vspace{-.15cm}
\subsection{Phase 2: Column-Generation Approach}
\label{phase2cg}

Since our definition of configurations ($\ConfigIdx$) has been simplified, CG becomes linear and faster.

\subsubsection{Restricted Master Problem (\textbf{RMP}) }
\label{masterProb}

%
%

\vspace{0.2cm}
\noindent
\textbf{Variables:}
\begin{itemize}
\item $z_{\ConfigIdx} =1$ if configuration $\ConfigIdx$ is selected; 0 otherwise.
\item $\xvic =1$ if $i$th function of $\ServiceChainIdx$ is located in $v$; 0 otherwise.
\item $y^{\text{first(\ServiceChainIdx)}, sd}_{\ell} =1$ if $\ell$ is on path from $v_s$ to location of first VNF in $\ServiceChainIdx$; 0 otherwise.
\item $y^{\text{last(\ServiceChainIdx)}, sd}_{\ell} =1$ if $\ell$ is on path from location of last VNF in $\ServiceChainIdx$ to $v_d$; 0 otherwise.
\item $h_v =1$ if $v$ is used as a location for a VNF; 0 otherwise.
\end{itemize}

\vspace{0.2cm}
\noindent
\textbf{Objective:} Minimize bandwidth consumed:
\begin{multline} \min \quad 
     \sum\limits_{\ConfigIdx \in \ConfigSet}  \>
     \underbrace{ 
     \overbrace{ \left(   \sum\limits_{(s,d) \in \mathcal{SD}}   \> D^c_{sd}  \right) }^{\text{Overall traffic using } \ServiceChainIdx} 
     \overbrace{ \left( \sum\limits_{\ell \in L} \> \sum\limits_{i \in I} b_{i \ell}^{\ConfigIdx} \right)}^{ \substack{\text{Number of links} \\  \text{ in the route of } \ServiceChainIdx}}
                   }_{\textsc{cost}_{\ConfigIdx}} 
                   \zRMP  + \\
     \sum\limits_{\ServiceChainIdx \in \ServiceChainSet} \> 
     \sum\limits_{\ell \in L} \> \sum\limits_{(s,d) \in \mathcal{SD}}   \> D^c_{sd}  \left( y^{f_1(\ServiceChainIdx), sd}_{\ell} + y^{f_{n_c}(\ServiceChainIdx), sd}_{\ell} \right)    
\end{multline}

Total bandwidth consumed in placing multiple SCs depends on configuration  $\ConfigIdx$  selected for each SC  $\ServiceChainIdx$. Each $\ConfigIdx$ for $\ServiceChainIdx$ locates VNFs of $\ServiceChainIdx$ and gives the route to traverse these VNF locations. So, bandwidth consumed when going from $v_s$ to $v_d$ and traversing the SC depends on selected $\ConfigIdx$.  


\noindent
\textbf{Constraints}:
\begin{alignat}{2}
 & \sum\limits_{{\ConfigIdx} \in \ConfigSet_\ServiceChainIdx} z_{\ConfigIdx} = 1  
 && \hspace*{-4.cm} \ServiceChainIdx \in \ServiceChainSet \label{eq:single_config_per_service_chain} \\
& \sum\limits_{\SFCidx \in \SFCset} \> \sum\limits_{\ConfigIdx \in  \ConfigSet_{\SFCidx}} 
     \> \sum\limits_{(v_s, v_d) \in \mathcal{SD}}  D_{sd}^c 
     \> (\sum\limits_{i=1}^{n_c}  a_{v i}^{\ConfigIdx} \deltafic  \ncoref) \> \zRMP \leq \ncore 
&& \myspace \nonumber \\
& &&  \hspace*{-4.cm} v \in V^{\NFV} \label{eq:capa_cores} \\
&   \sum\limits_{\ServiceChainIdx \in \ServiceChainSet} \>  \sum\limits_{(v_s, v_d) \in \mathcal{SD}}  \> D^{\ServiceChainIdx}_{sd}  && \nonumber \\
&\qquad  \left(    y^{f_1(\ServiceChainIdx), sd}_{\ell} 
        + y^{f_{n_c}(\ServiceChainIdx), sd}_{\ell}     
        + \sum\limits_{{\ConfigIdx} \in \ConfigSet_{\ServiceChainIdx}} \> \sum\limits_{i =1}^{n_{\ServiceChainIdx} - 1}  b_{i \ell}^{\ConfigIdx} \>    
   \zRMP \right)   && \nonumber \\
& \qquad \qquad \qquad \qquad 
  \leq \textsc{cap}_{\ell}  
&& \hspace*{-2.cm}  \ell \in L \label{eq:capacity} \\
& \sum\limits_{\ConfigIdx \in \ConfigSet_{\ServiceChainIdx}}  a_{v i}^{\ConfigIdx}  \zRMP =  \xvic  
&&   \hspace*{-4.cm} f_i \in F(c), \ServiceChainIdx \in \ServiceChainSet, v \in V^{\NFV} \label{eq:a_x_consistent1} \\
&  M x_{vf} \geq \sum\limits_{\ServiceChainIdx \in \ServiceChainSet: f \in \ServiceChainIdx} \> \sum\limits_{i \in \{1,2,\dots, n_c\}: f_i = f}  \xvic    \geq x_{vf} 
 && \nonumber \\
& && \hspace*{-3.cm} v \in V^{\NFV}, f_i \in F \label{eq:f_on_v} \\
& M h_v \geq \sum\limits_{f \in F} x_{vf} \geq h_v && \hspace*{-3.cm} v \in V^{\NFV} \label{eq:vnf_location} \\
& \sum\limits_{v \in V^{\NFV}} h_v \leq K \label{eq:k_nfv_nodes}
\end{alignat}  

Constraints \eqref{eq:single_config_per_service_chain} guarantee that we select exactly one $\ConfigIdx$ for SC  $\ServiceChainIdx$  
and force $\ServiceChainIdx$ to have a single instance. 
Each $\ConfigIdx$ is associated with a set of $a^{\ConfigIdx}_{vi}$ (from \textbf{PP\_SC($ \ServiceChainIdx$)}) required to be consistent with $\xvic$ in \textbf{RMP}, which is resolved by Eqs. \eqref{eq:a_x_consistent1}.

Constraints \eqref{eq:capa_cores} ensure that each NFV node has a sufficient number of CPU cores for hosting $f$. 
Eq. \eqref{eq:capacity} enforces link-capacity constraints for the complete route for SC  $\ServiceChainIdx$  from $v_s$ to $v_d$  for all $(v_s,v_d) \in \mathcal{SD} : D^{c}_{sd} > 0$).

Eq. \eqref{eq:f_on_v} keeps track of VNF replicas. Eq. \eqref{eq:vnf_location} keeps track of NFV nodes used for hosting VNFs while Eq. \eqref{eq:k_nfv_nodes} enforces the number of NFV nodes allowed to host VNFs.

%
%

\begin{alignat}{2}     
& \text{\textbf{Route from} } v_s \text{ \textbf{to first function location:}} \nonumber \\
& \sum\limits_{\ell \in \omega^+{(v_s)}} y^{f_1(\ServiceChainIdx), sd}_{\ell} = 1 - \xone
&& \myspace  \ServiceChainIdx \in \ServiceChainSet, \nonumber \\
& &&  \hspace*{-3.5cm}  (v_s,v_d) \in \mathcal{SD}: D_{sd}^{\ServiceChainIdx} > 0 \label{eq:link_from_source_to_ingress} \\
& \sum\limits_{\ell \in \omega^-{(v)}} y^{f_1(\ServiceChainIdx), sd}_{\ell}  \geq \xvone
&&  \hspace*{-2.cm}  \ServiceChainIdx \in \ServiceChainSet, \nonumber \\
& && \hspace*{-5.cm}  (v_s,v_d) \in \mathcal{SD}: D_{sd}^{\ServiceChainIdx} > 0, v \in V^{\NFV} \setminus \{ v_s \} \label{eq:to_ensure_NFV_1_placement} \\
&     \sum\limits_{\ell \in \omega^+{(v)}} y^{f_1(\ServiceChainIdx), sd}_{\ell}
     -  \sum\limits_{\ell \in \omega^-{(v)}} y^{f_1(\ServiceChainIdx), sd}_{\ell} 
     = - \xvone
&& \nonumber \\
& && \hspace*{-6.cm}  \ServiceChainIdx \in \ServiceChainSet, 
            (v_s,v_d) \in \mathcal{SD}: D_{sd}^{\ServiceChainIdx} > 0, v \in V^{\NFV} \setminus \{ v_s \}        \label{eq:places_NFV} \\
&     \sum\limits_{\ell \in \omega^+{(v)}} y^{f_1(\ServiceChainIdx), sd}_{\ell}
     -  \sum\limits_{\ell \in \omega^-{(v)}} y^{f_1(\ServiceChainIdx), sd}_{\ell}
     = 0
      && \nonumber \\
      & && \hspace*{-7.cm}  \ServiceChainIdx \in \ServiceChainSet, (v_s,v_d) \in \mathcal{SD}: D_{sd}^{\ServiceChainIdx} > 0, v \in V \setminus (V^{\NFV} \cup \{ v_s \})         \label{eq:places_non_NFV} 
\end{alignat} 

We assume that an unique route exists from $v_s$ to first VNF location. This is imposed by selecting exactly one outgoing link from $v_s$ unless first VNF is located at 
$v_s$. We account for these scenarios using Eq. \eqref{eq:link_from_source_to_ingress}. To find the route from $v_s$ to first VNF, flow conservation needs to be enforced at the intermediate nodes which may or may not have NFV support. Eqs. \eqref{eq:places_NFV} and \eqref{eq:places_non_NFV} enforces flow-conservation constraints at nodes with and without NFV support, respectively.

We can enforce same functionality as Eqs. \eqref{eq:link_from_source_to_ingress},  \eqref{eq:places_NFV}, \eqref{eq:places_non_NFV}, and \eqref{eq:to_ensure_NFV_1_placement},  on route from location of last VNF to $v_d$. For the interested reader, similar details are provided in \cite{gupta_icc_arxiv}.

\subsubsection{Pricing Problem}
\label{pricingProb}

Mapping configurations for each SC  $\ServiceChainIdx$  ($\ServiceChainIdx \in \ServiceChainSet$) corresponds to the solution of pricing problems. The number of pricing problems to be solved equals the sum of the number of SC {\SCcopies} to be deployed. 

Pricing problem \textbf{PP\_SC($ \ServiceChainIdx$)} generates: 
\textit{(i)} A set of locations for VNFs of  $\ServiceChainIdx$; and
\textit{(ii)} a sequence of paths from the location of VNF $f_i$ to the location of VNF $f_{i+1}$, for $i = 1, 2, \dots, n_\ServiceChainIdx - 1$ for chain $\ServiceChainIdx$. Each solution that is generated by \textbf{PP\_SC($\ServiceChainIdx$)} with a negative reduced cost leads to a new potential $\ConfigIdx$ for  $\ServiceChainIdx$ of interest. Please see \cite{gupta_icc_arxiv} for further details.

Let $u_{c}^{\eqref{eq:single_config_per_service_chain}} \lesseqgtr 0, u^{\eqref{eq:capa_cores}}_{v} \geq 0,$ and, $u_{vf}^{\eqref{eq:a_x_consistent1}} \geq 0$ be values of dual variables associated with constraints \eqref{eq:single_config_per_service_chain}, \eqref{eq:capa_cores}, \eqref{eq:a_x_consistent1}, respectively.

\vspace{.2cm} 

\noindent
\textbf{Variables}: 
\begin{itemize}
\item $a_{vi}$ = 1 if $i$th function $f_i$  of $\ServiceChainIdx$ is located in $v \in V^{\NFV}$; 0 otherwise. 
\item $ b_{i\ell}$ = 1 if $\ell$ is on the path from location of  $f_i$ to location of $f_{i+1}$; 0 otherwise.
\end{itemize}

\vspace{.2cm}

\noindent
\textbf{Objective:} Minimize reduced cost of variable $z_{\ConfigIdx}$: 
\begin{multline}
\text{[\textbf{PP\_SC($\ServiceChainIdx$)}]} \qquad 
 \redcost_{\ConfigIdx} = \cost_{\ConfigIdx}
- u^{\eqref{eq:single_config_per_service_chain}} \\
+ \sum\limits_{v \in V^{\NFV}} \> u^{\eqref{eq:capa_cores}}_{v}  \>{\sum\limits_{(v_s, v_d) \in \SD} }D_{sd}^{\ServiceChainIdx} \> 
   \sum\limits_{i =1}^{n_c}   \ncorefi   a_{v i}  \\
+ \sum\limits_{\ell \in L} \sum\limits_{(v_s, v_d) \in \SD} u^{\eqref{eq:capacity}}_{\ell} D_{sd}^c 
                  \sum\limits_{i =1}^{n_{\ServiceChainIdx} - 1} b_{i\ell} 
-   \sum\limits_{i =1}^{n_{\ServiceChainIdx}} \> \sum\limits_{v \in V^{\NFV}} u_{vci}^{\eqref{eq:a_x_consistent1}} a_{vi}.              
\end{multline}

\noindent
where $\redcost$ value indicates whether an optimal $\ConfigIdx$ for $\ServiceChainIdx$ has been found. A non-negative value of $\redcost$ indicates optimality for our model.

\vspace{.1cm}
\noindent 
\textbf{Constraints:}
\begin{alignat}{2}
& \sum\limits_{(v_s, v_d) \in \SD} \> D^{\ServiceChainIdx}_{sd} \> \sum\limits_{i=1}^{n_c}  \ncoref \deltafic a_{vi} \leq \ncore && \nonumber\\
& &&    v \in V^{\NFV} \label{eq:PP_node_capacity} \\
& \sum\limits_{(v_s,v_d) \in \SD} D^{c}_{sd} \sum\limits_{i=1}^{n_c - 1} b_{i \ell} \leq \CAP_{\ell}    \quad 
&&     \ell \in L \label{eq:PP_link_capacity} 
\end{alignat} 

Eqs. \eqref{eq:PP_node_capacity} and \eqref{eq:PP_link_capacity} are compute resource and capacity constraints, similar to those in \textbf{RMP} and are linear.
The rest of the equations are the same as Eqs. \eqref{eq2:each_function} to \eqref{eq2:flow} in Section \ref{colGenPP} and perform the same function.


\subsubsection{Solution Scheme}
\label{solSchemeTwoPm}
Same as in Section \ref{solScheme}.

\begin{figure}
  \centering
  \begin{tabular}{cc}
     \subfloat[][4 node]{\label{fig:a}\includegraphics[width=.2\textwidth, scale=1]{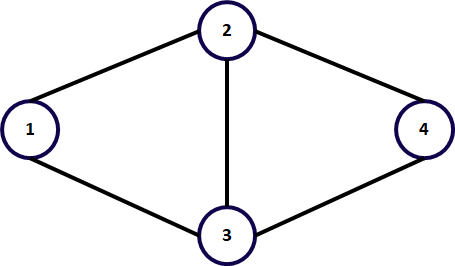}} 
     & \subfloat[][5 node]{\label{fig:b}\includegraphics[width=.2\textwidth, scale=1]{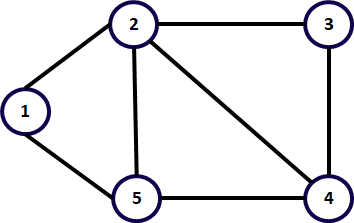}} \\
    \multicolumn{2}{c}{ \subfloat[][6 node]{\label{fig:c}\includegraphics[width=.25\textwidth, scale=1]{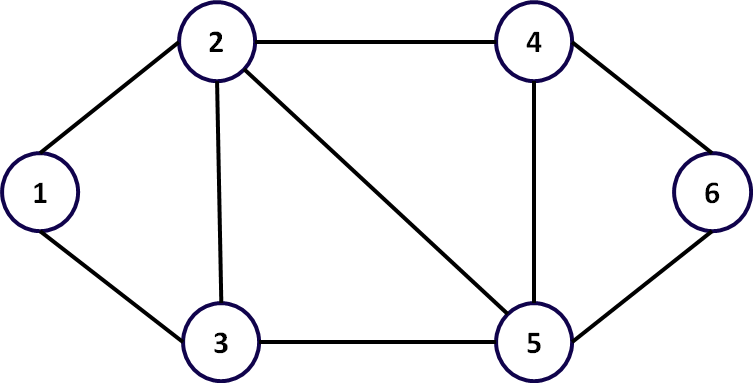}} }
  \end{tabular}
  \caption{Network topologies.}
  \label{fig:nettopo}
\end{figure}

\begin{figure}[htb]	
  	\centering
    	\includegraphics[width=0.35\textwidth]{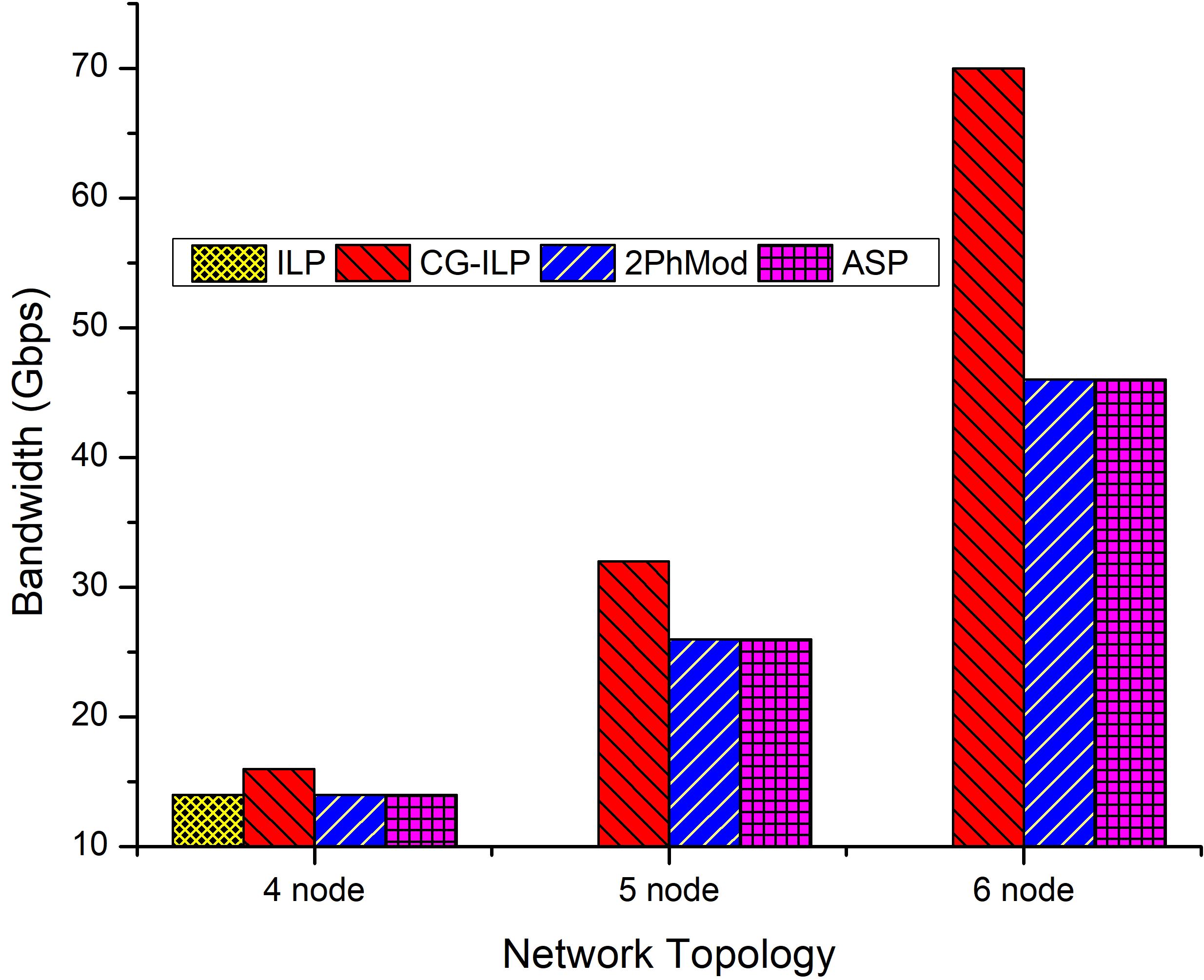}
    \caption{Comparison of bandwidth used.}
    \label{fig:bw_sol_methods}
\end{figure}

\begin{figure}[htb]	
  	\centering
    	\includegraphics[width=0.35\textwidth]{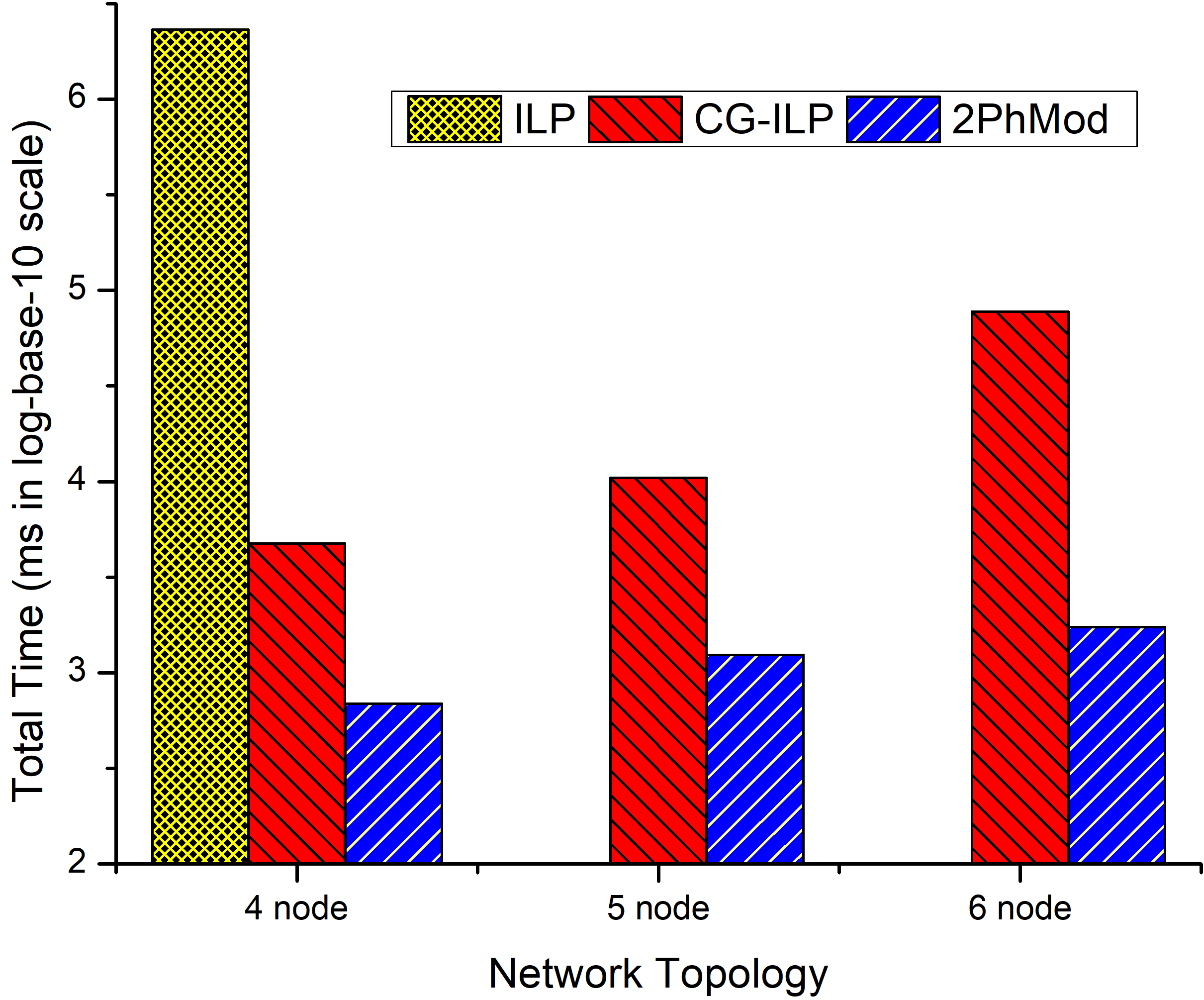}
    \caption{Comparison of total time.}
    \label{fig:tt_sol_methods}
\end{figure}

\begin{figure*}
  \centering
  \begin{tabular}{cc}
     \subfloat[][NSFNET K=14]{\label{fig:aa}\includegraphics[width=.5\textwidth, scale=1]{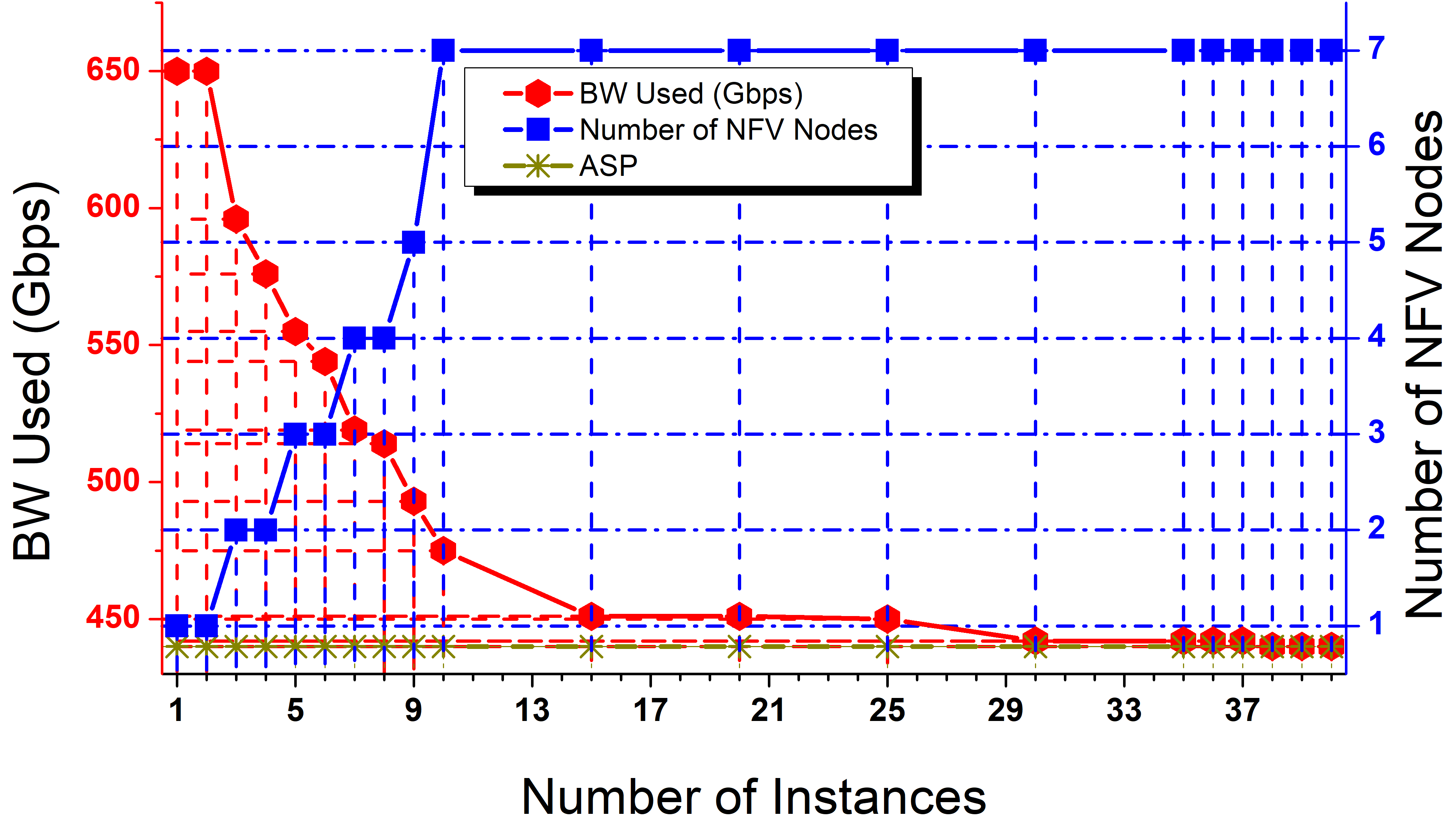}}
    & \subfloat[][NSFNET K=1,2,3,4,5,14]{\label{fig:bb}\includegraphics[width=.45\textwidth, scale=1]{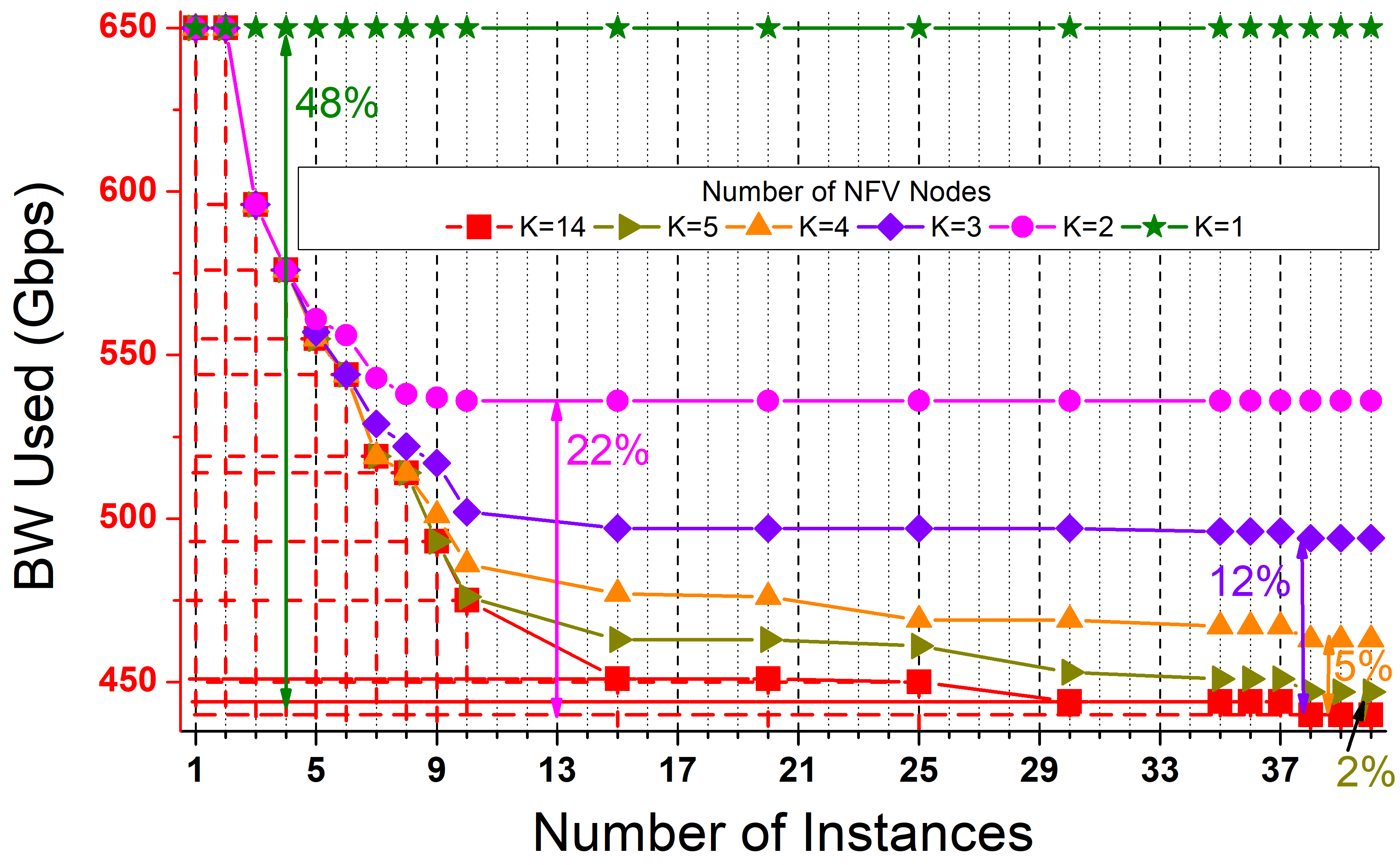}}
    \\ \\
     \subfloat[][COST239 K=11]{\label{fig:cc}\includegraphics[width=.5\textwidth, scale=1]{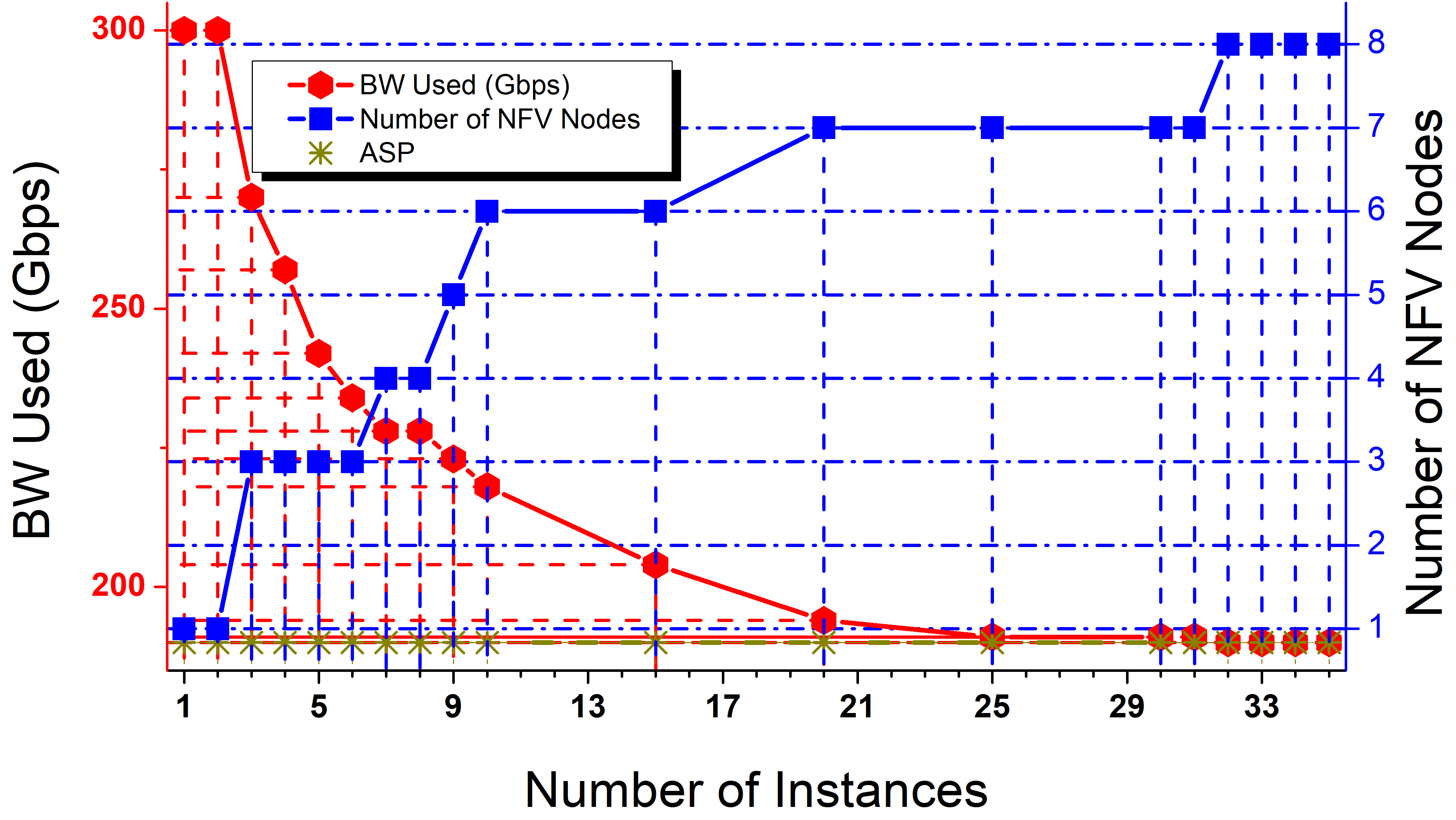}}
    & \subfloat[][COST239 K=1,2,3,4,5,11]{\label{fig:d}\includegraphics[width=.45\textwidth, scale=1]{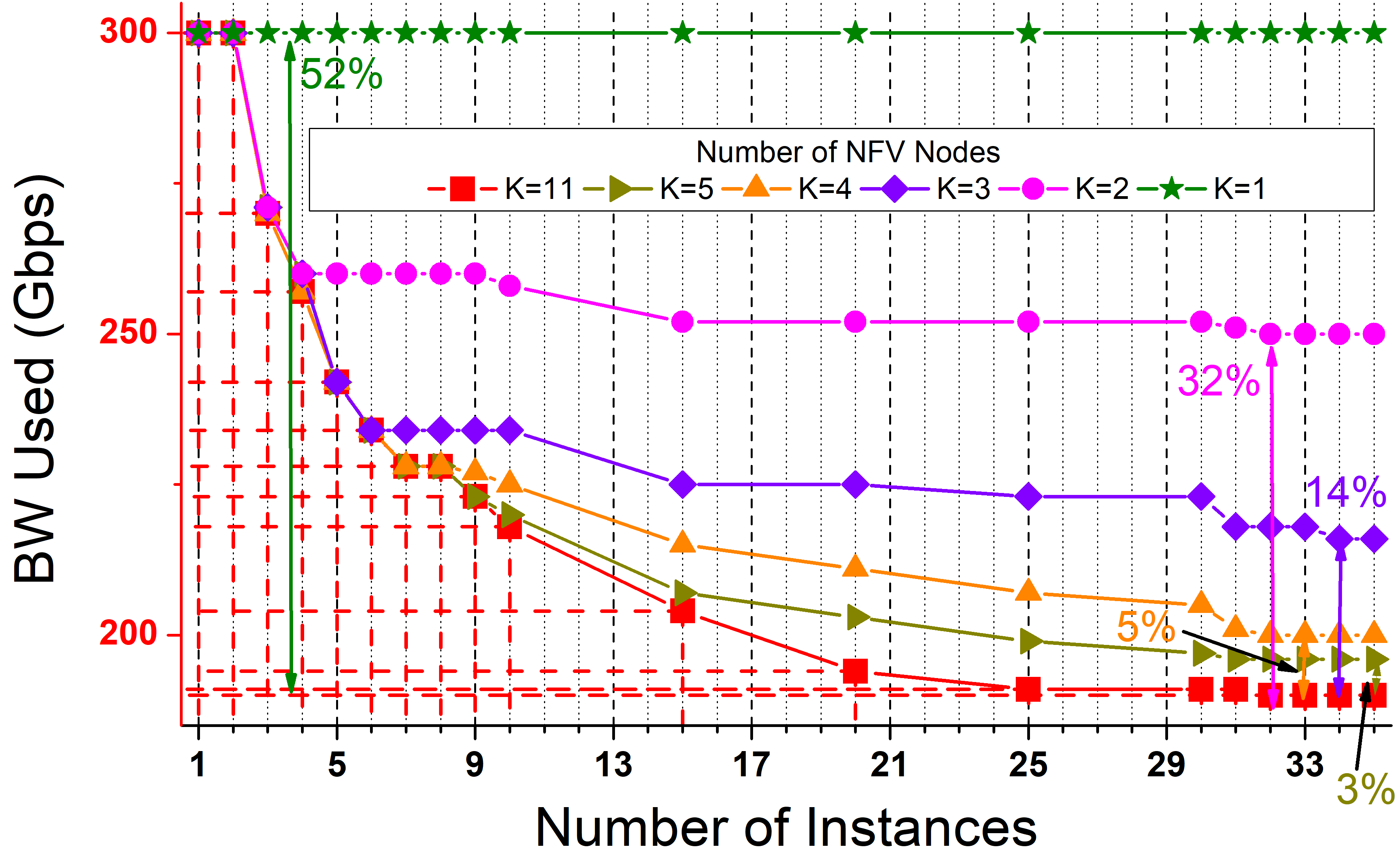}}  
  \end{tabular}
  \caption{Bandwidth vs. number of NFV nodes in NSFNET and COST239 networks.}
  \label{results}
\end{figure*}

\section{Comparison of Solution Approaches}

To benchmark our solution approaches ILP, CG-ILP, and Two-Phase model, we use the All Shortest Path (ASP) calculation. ASP assumes that, in the best possible scenario, all traffic flows requiring a SC $c$ will have a SC instance deployed on their shortest path. Thus, total bandwidth used will be equal to all traffic flows taking a shortest path.

Fig. \ref{fig:bw_sol_methods} compares bandwidth consumption of our three approaches across three network topologies shown in Fig. \ref{fig:nettopo} for a single SC $c$ deployment. We consider all-to-all traffic flows in each network and allow $I_c$ instances for all solution approaches. ILP is shown to be as good as ASP for four-node networks; however, it does not scale for larger networks because of pre-computation of all possible configurations. CG-ILP does not provide optimal solutions because of $\varepsilon$-optimality gap (difference between ILP and LP values). Two-Phase model performs as well as ASP for all topologies.

Fig. \ref{fig:tt_sol_methods} shows total time taken by various approaches. Note that Two-Phase Model scales best across all topologies. 

\section{Illustrative Numerical Examples}
\label{sim_examples}

\subsection{Single Service Chain Scenario}
\label{single_sc_results}

\begin{table}
 \centering
 \begin{tabular}{|c|c|c|} 
 \hline
 Service Chain & Chained VNFs & \%traffic \\ [0.5ex] 
 \hline
 Web Service & NAT-FW-TM-WOC-IDPS & 18.2\% \\ 
 VoIP & NAT-FW-TM-FW-NAT & 11.8\% \\
 Video Streaming & NAT-FW-TM-VOC-IDPS & 69.8\% \\ 
 Online Gaming & NAT-FW-VOC-WOC-IDPS & 0.2\% \\
 \hline 
 \end{tabular}
 \caption {Service Chain Requirements \cite{sc_detail}; Network Address Translator (NAT), Firewall (FW), Traffic Shaper (TM), WAN Optimization Controller (WOC), Intrusion Detection and Prevention System (IDPS), Video Optimization Controller (VOC).}
 \label{table:sc_chain_req}
\end{table}

We first tested our two-phase optimization process on a 14-node NSFNET WAN topology \cite{gupta_icc_arxiv} with a complete traffic matrix, i.e., with traffic flows between all node pairs, assuming all nodes can be made NFV nodes. The link capacities are sufficient to support all flows. Each traffic flow is 1 Gbps and demands the same 5 VNF service chain (SC) for video streaming, as shown in Table \ref{table:sc_chain_req}. Compute resource (CPU) at each node is sufficient to support traffic demand, which helps in determining the optimal location to deploy CPU cores and number of CPU cores at each location. The second run of the model is on an 11-node COST239 WAN topology \cite{farhan_cost239} under the same specifications as above.    

Figure \ref{results}\subref{fig:a} shows the bandwidth consumption as the number of SC {\SCcopies} increases. 
Here, we allow all nodes (K=14) to host VNFs. 
We find that, as number of deployed SC {\SCcopies} increases, bandwidth consumption decreases. With a higher number of {\SCcopies}, more groups of traffic node pairs are able to take short paths. We see that, at 38 {\SCcopies}, we achieve minimum possible bandwidth consumption, meaning all traffic flows are taking the shortest path. 
Note that number of traffic node pairs in the network is 182, requiring apriori upto 182 different {\SCcopies} (solving the problem for 182 {\SCcopies} would be equivalent to obtaining a solution with existing models as in \cite{place_vnf_secci}\cite{vnf_placement_barcellos_gaspary}\cite{orc_vnf_boutaba}). Instead, our approach, with only 38 {\SCcopies}, achieves minimum bandwidth consumption. This is important as an operator may deploy multiple SCs and manage multiple {\SCcopies} per SC including routing flows to a particular SC {\SCcopy}. So, a lower number of {\SCcopies} will lower the orchestration overhead for the operator. 

\begin{figure}[htb]	
  	\centering
    	\includegraphics[width=0.45\textwidth]{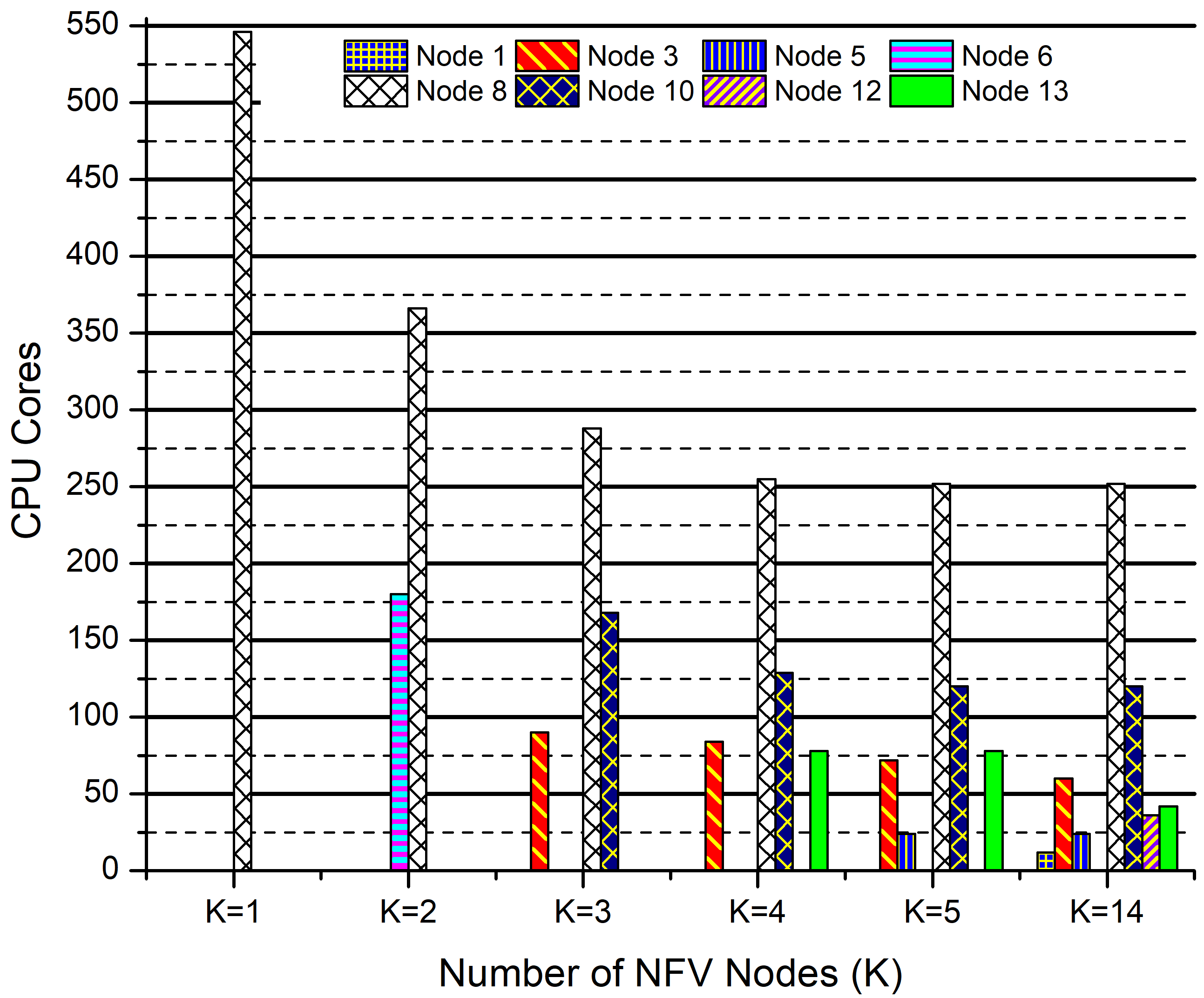}
    \caption{CPU core distribution across K (NSFNET).}
    \label{fig:core_distribution}
\end{figure}

On the other hand, number of NFV nodes increases as number of SC {\SCcopies} increases. Indeed, as SC mappings become more distributed, more nodes are being used for hosting virtual functions. In Fig. \ref{results}\subref{fig:aa}, 11 nodes are NFV enabled for 38 different SC mappings. 
For a network operator, capital expenditure in making 11 out of 14 nodes capable of hosting VNFs is very high. 
So, operators may want to minimize the number of NFV nodes while also trying to reduce bandwidth consumption by using multiple SC mapping {\SCcopies}. This led us to explore how the bandwidth consumption varies when the numbers of NFV nodes are limited.  

\begin{figure}[htb]	
  	\centering
    	\includegraphics[width=0.45\textwidth]{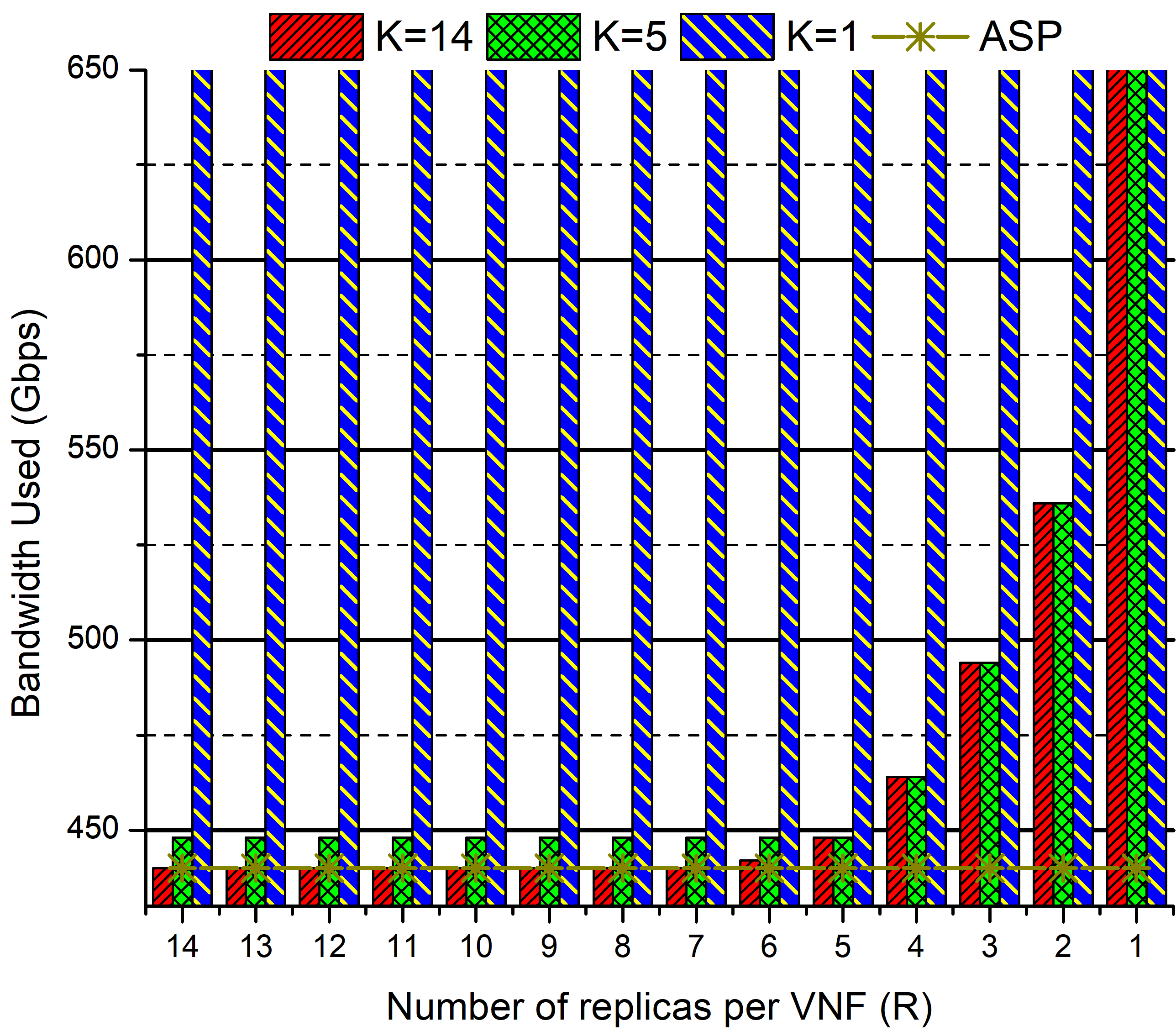}
    \caption{Bandwidth used across R (NSFNET).}
    \label{fig:vnf_replica_single_sc}
\end{figure}

Figure \ref{results}\subref{fig:b} shows bandwidth consumption for SC mapping {\SCcopies} for various $K$ values. When $K=1$, all traffic flows have to traverse the one node in the network; hence, number of {\SCcopies} does not affect bandwidth consumption. At $K=2$, deploying more than 10 {\SCcopies} does not improve bandwidth utilization. For $K=3$ and 35 {\SCcopies}, we are able to achieve close to 10\% of the minimum bandwidth utilization. Similarly, at $K=4$, we reach within 5\% of the optimal bandwidth consumption. Bandwidth consumption comes to within 2\% of the optimal when K=5 and 38 {\SCcopies}. Thus, we can achieve near-optimal bandwidth consumption by a using a small number of {\SCcopies} and nodes.

\begin{figure}[tb]	
  	\centering
    	\includegraphics[width=0.45\textwidth]{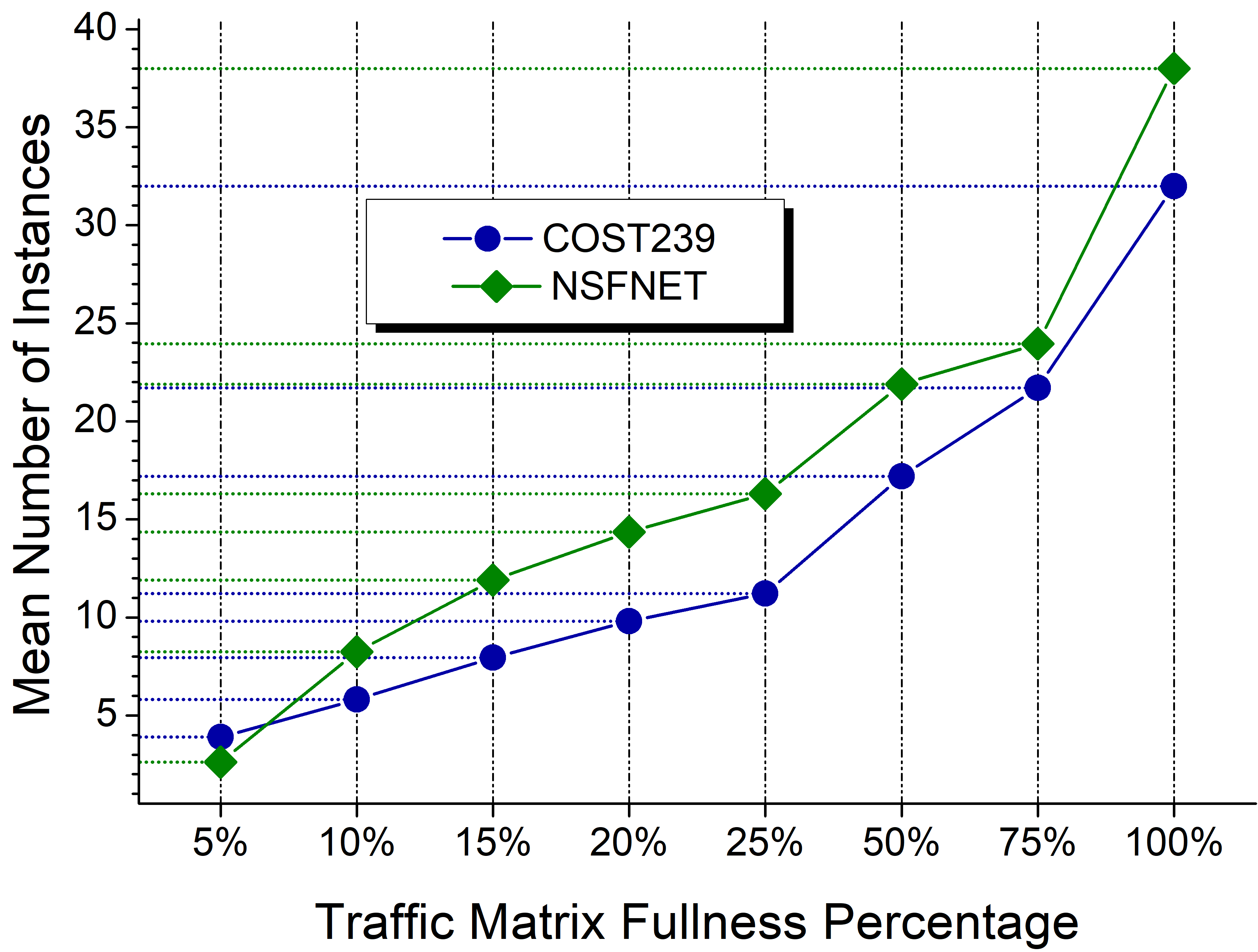}
    \caption{Mean number of SC instances across varying number of traffic flows.}
    \label{fig:mean_count}
\end{figure}

Figures \ref{results}\subref{fig:cc} and \ref{results}\subref{fig:d} corroborate our findings in Figs. \ref{results}\subref{fig:aa} and \ref{results}\subref{fig:bb} over COST239 network.

\begin{figure}[htb]	
  	\centering
    	\includegraphics[width=0.45\textwidth]{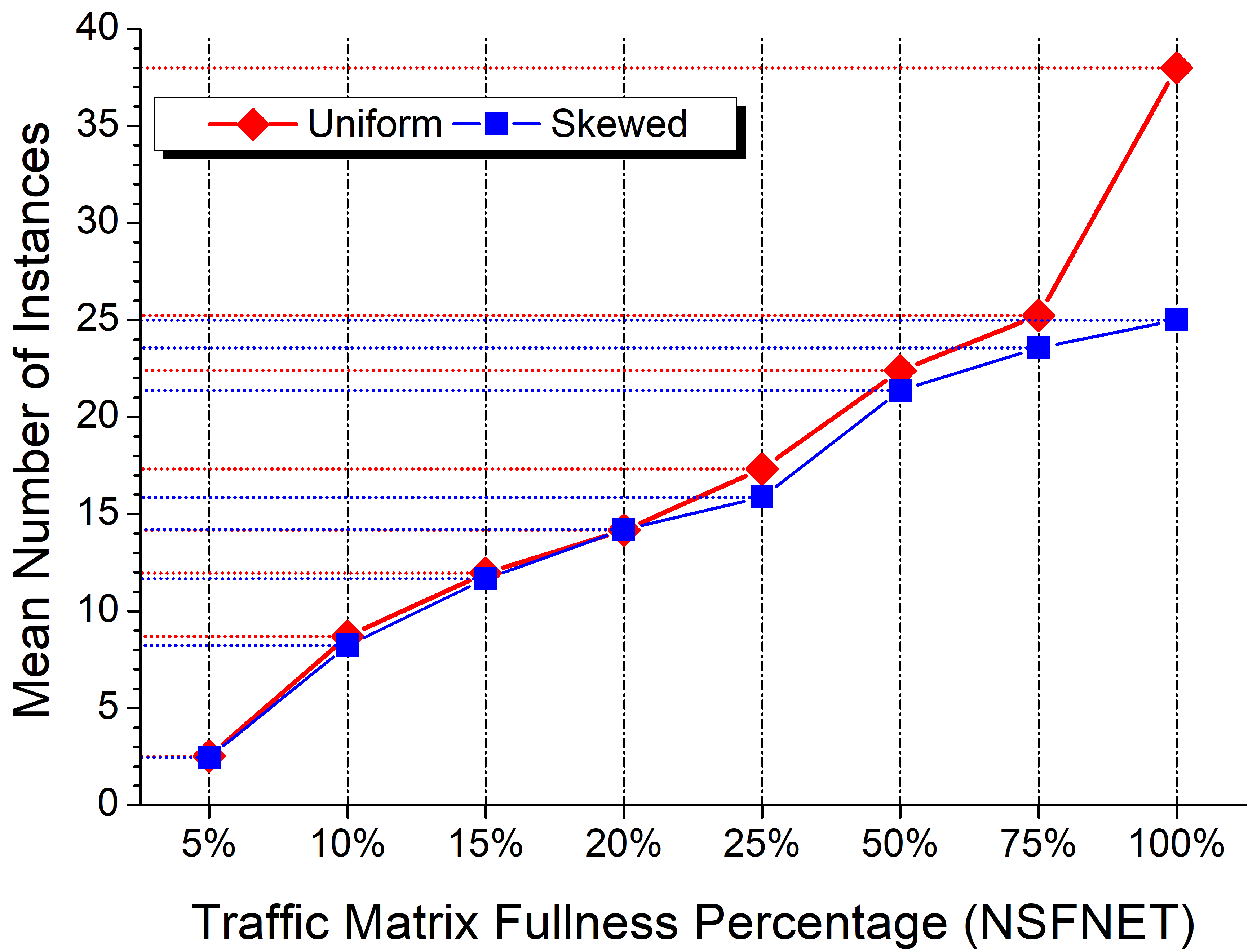}
    \caption{Mean number of SC instances for uniform and skewed traffic.}
    \label{fig:compare_mean_count}
\end{figure}

\begin{figure*}
  \centering
  \begin{tabular}{cc}
     \subfloat[][NSFNET K=14]{\label{fig:aa}\includegraphics[width=.48\textwidth, scale=1]{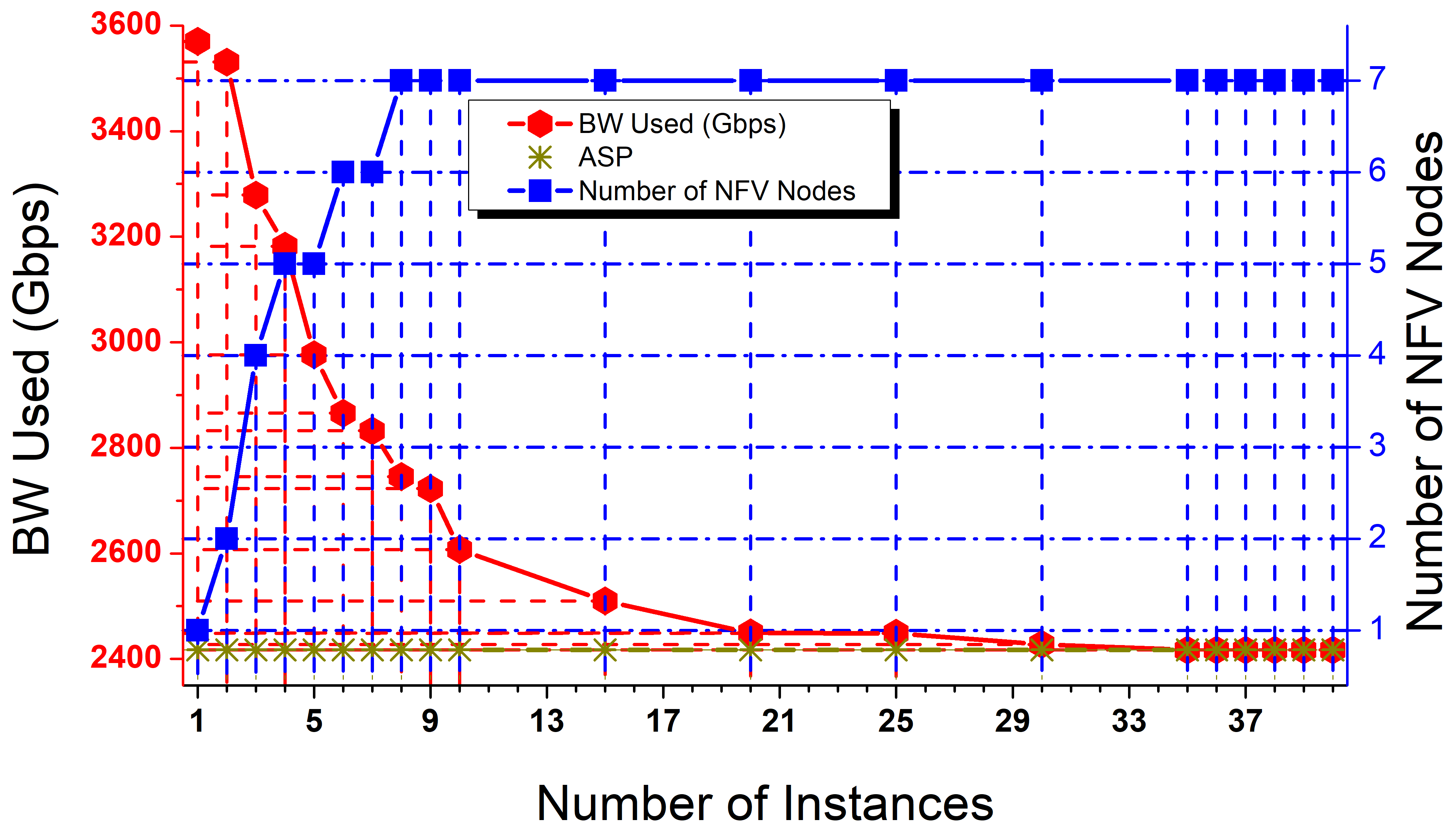}}
    & \subfloat[][COST239 K=11]{\label{fig:bb}\includegraphics[width=.48\textwidth, scale=1]{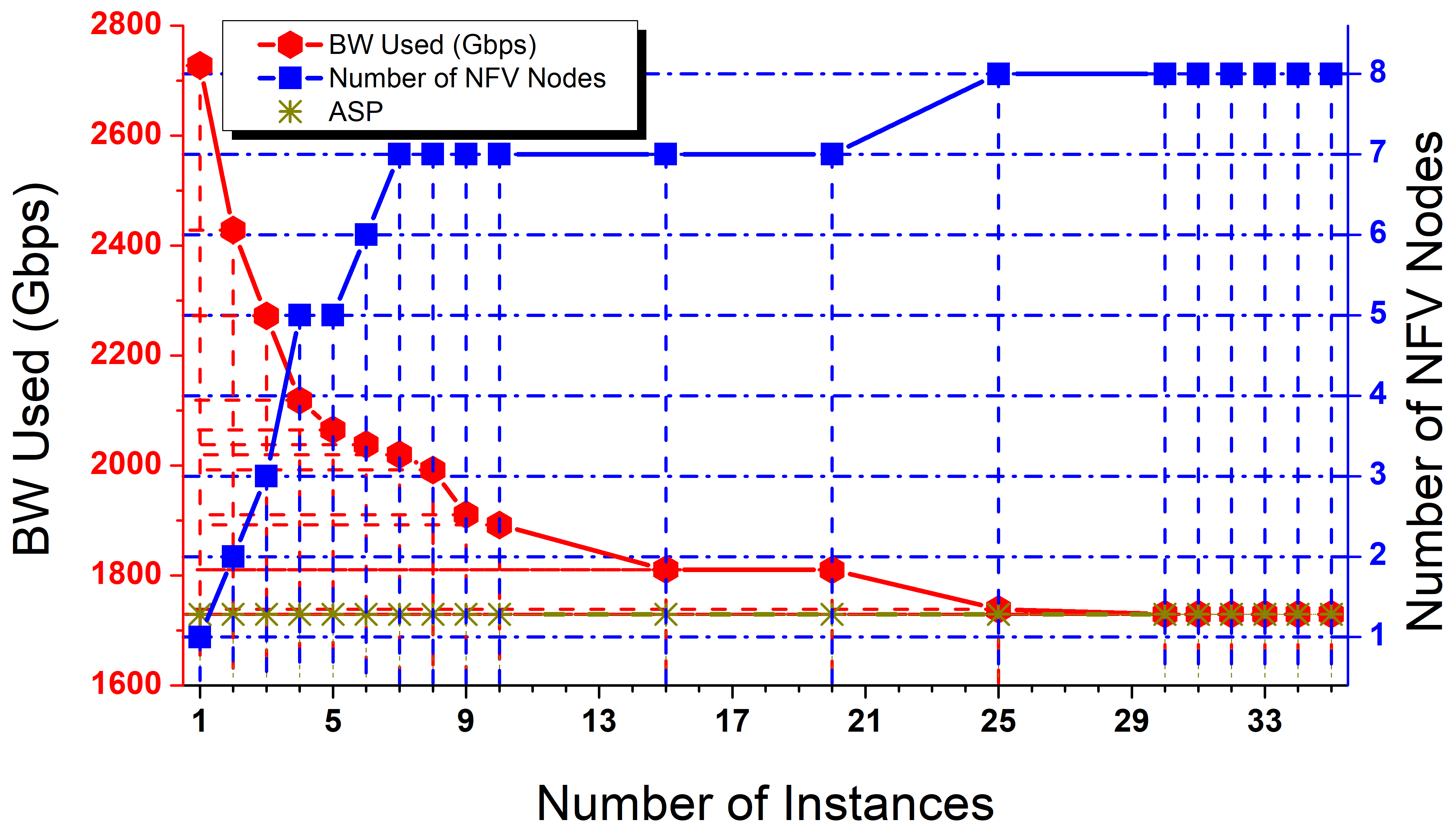}}   
  \end{tabular}
  \caption{Bandwidth vs. number of NFV Nodes in NSFNET and COST239 when deploying all service chains in Table \ref{table:sc_chain_req}.}
  \label{fig:results_all_sc}
\end{figure*}

Figure \ref{fig:core_distribution} shows amount of compute resources (CPU cores) and their location for different values of K, given that we know optimal number of SC {\SCcopies} required. CPU cores used by each VNF depends on VNF type and throughput required as shown in Table \ref{table:cpu_core_values}. At $K=1$, node 8 will be selected for deploying CPU cores. When $K=2$, the best location for deploying CPU resources are nodes 8 and 6. Note that the traffic remains the same across $K=1$ and $K=2$, so the total amount of CPU cores used remain the same. However, when $K=2$, two nodes get selected as it reduces the total bandwidth consumption, and the distribution of CPU cores happens across nodes. We find that more nodes are selected for $K=3,4,5,14,$ and the compute resources become more distributed. At K=14, we find that only 7 nodes are used to host CPU cores, which means we need to have at max 7 NFV nodes to achieve $ASP$ bandwidth consumption.

\begin{table}
\begin{center}
 \renewcommand{\arraystretch}{1.5}
 \begin{tabular}{| >{\centering\arraybackslash}m{0.7in} | >{\centering\arraybackslash}m{0.4in} | >{\centering\arraybackslash}m{0.4in} | >{\centering\arraybackslash}m{0.6in} |}
 \hline
  \multirow{2}{*}{Application}
  &\multicolumn{3}{|c|}{Throughput}\\
   \cline{2-4} 
  & 1 Gbps & 5 Gbps & 10 Gbps \\ [0.5ex]
 \hline
 NAT & 1 CPU & 1 CPU & 2 CPUs \\ [0.5ex]
 \hline
 IPsec VPN & 1 CPU & 2 CPUs & 4 CPUs \\ [0.5ex]
 \hline
 Traffic Shaper & 1 CPU &  8 CPUs &  16 CPUs \\ [0.5ex]
 \hline
 \end{tabular}
 \caption{VNF requirements as per throughput \cite{cisco_vnf}.}
\label{table:cpu_core_values}
\end{center}
\end{table}

In the above results, we determine the number of SC {\SCcopies} required for each $K$ to get minimum bandwidth consumption. We define this count of SC {\SCcopies} to be optimal. Now, given this optimal number of SC {\SCcopies}, we want to observe the effect different number of replicas of VNFs ($R$) has on different $K$ values. Here, $R=14$ means all VNFs in the SC are allowed 14 replicas. Fig. \ref{fig:vnf_replica_single_sc} compares bandwidth used in NSFNET (when $K=1,5,14$) when different $R$ are allowed. We find that when $K=5,R=5$ our bandwidth consumption is close to $ASP$, implying that we require a small number of $K$ and $R$.

Figure \ref{results}\subref{fig:a} shows the number of SC {\SCcopies} required to achieve $ASP$ bandwidth consumption when there are traffic flows between all traffic pairs. We call this a 100\% traffic matrix fullness, i.e., all entries in the traffic matrix have been filled. Fig. \ref{fig:mean_count} shows the mean number of SC {\SCcopies} required to reach $ASP$ bandwidth consumption across different traffic matrix fullness percentages (percentage of entries that are filled in the traffic matrix) under the same traffic load. We find that the mean number of SC {\SCcopies} required to reach $ASP$ bandwidth consumption increases as the traffic matrix fullness percentage increases for both COST239 and NSFNET.

All results until now assumed an uniform traffic distribution. However, traffic can be skewed. So, we skew the traffic load based on \cite{skewed_traffic} (skewed based on population size of the nodes) for varying number of traffic flows (traffic matrix fullness percentage) and display the number of SC {\SCcopies} required to achieve $ASP$ bandwidth consumption. We compare $SPTG$ performance for uniform and skewed traffic in Fig. \ref{fig:compare_mean_count}. We find that $SPTG$ can achieve $ASP$ bandwidth consumption for skewed traffic distribution for lower number of SC instances, especially as number of traffic flows increase.

\subsection{Multiple Service Chain Scenario}
\label{multiple_sc_results}

In the previous subsection, we performed simulations where all traffic flows require the same service chain, i.e., all traffic requires the same VNFs. However, when traffic requires different service chains, not all VNFs are required by all traffic requests and the conclusions on a single SC may not hold. Hence, it becomes important to analyze the effect of varying the number of allowed VNF replicas ($R$) to focus on the role of each VNF on bandwidth consumption separately. In this subsection, we jointly deploy the four service chains in Table \ref{table:sc_chain_req} for a total traffic load of 1 Tbps. The distribution of traffic across service chains follows realistic relative popularity of the four services (see last column in Table \ref{table:sc_chain_req}). All four service chains are requested by all traffic pairs in the network, i.e., all four service chains have 100\% traffic matrix fullness. 

Figure \ref{fig:results_all_sc}\subref{fig:aa} shows bandwidth consumption as SC {\SCcopies} increase for all SCs deployed in NSFNET. We find that 35 instances for each SC deployed is sufficient to achieve $ASP$ bandwidth consumption. Number of NFV nodes used also does not vary much from previous result of Fig. \ref{results}\subref{fig:a}. 

\begin{figure}[htb]	
  	\centering
    	\includegraphics[width=0.45\textwidth]{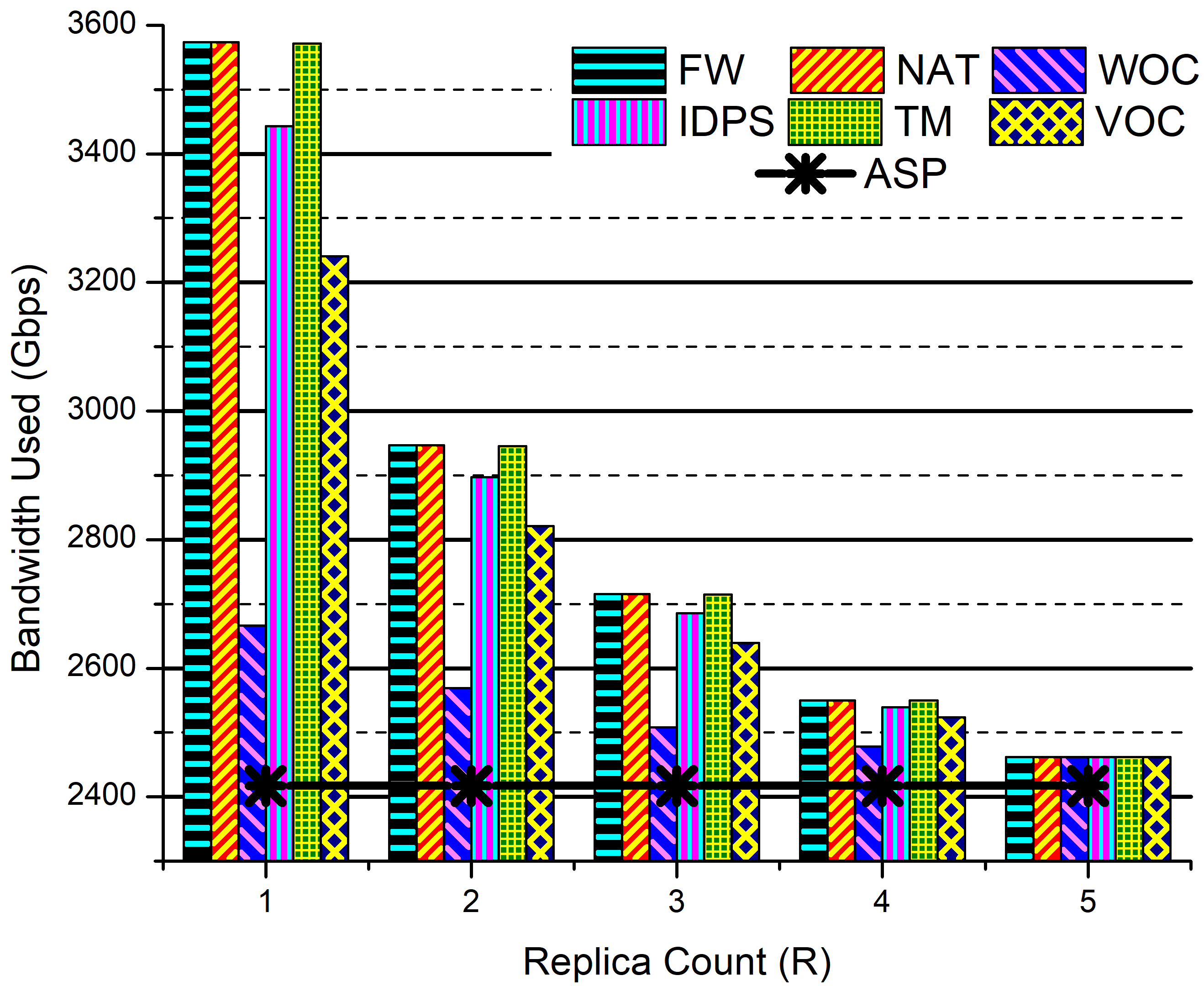}
    \caption{Bandwidth consumed for varying VNF replica (R) for $K=5$ (NSFNET).}
    \label{fig:bw_used_k5}
\end{figure}


We then analyze the effect of varying number of VNF replicas ($R$). 
Figure \ref{fig:bw_used_k5} shows the bandwidth consumed when $K=5$ and $R=1,2,3,4,5$. We reduce the number of replicas for a specified VNF to R while the remaining VNFs have R=K, i.e., R=5 here, for all the unspecified VNFs. So here, when $R=1$ for FW, that means the number of replicas allowed for FW is 1 while the other VNFs have replicas equal to K (here, $K=5$). We always see $R \leq K$ since a VNF can only be allowed to replicated in $K$ locations at maximum. Note that some VNFs like FW and NAT are shared across all service chains, while others like WOC and VOC are only shared across two service chains.  When $R=1$ for FW and NAT, highest bandwidth consumption is experienced as these VNFs are shared across all four service chains. Conversely, when $R=1$ for WOC, least bandwidth consumption is experienced since WOC is required by only 18.4\% of total traffic. When $R=2,3,4,$ bandwidth consumption reduces as $R$ increases, and decrease in bandwidth consumption across VNFs for each $R$ is seen to be dependent on amount of traffic requiring the VNF. This relative deference in bandwidth consumption between VNFs reduces as $R$ increases. This happens as $R$ becomes a less salient parameter for bandwidth consumption as $R$ approaches $K$. At $R=5$, bandwidth consumption is the same for all VNFs. This happens as when $R=5$ for FW, $R$ values for the unspecified VNFs are also 5. Each column when $R=5$ represents the same situation where all VNFs used are allowed 5 replicas. We find that when $K=5$, $R=5$ (when all VNFs have 5 replicas) is sufficient for achieving close to $ASP$ bandwidth consumption.


\vspace{-.2cm}
\subsection{Scalability}
\label{scale}

\begin{table}
 \centering
 \begin{tabular}{|c|c|c|c|} 
 \hline
 Network & Nodes & Links & Mean Time (s)\\ [0.5ex] 
 \hline
 COST239 & 11 & 44 & 12.1\\ 
 NSFNET & 14 & 40 & 14.2\\
 EON\cite{eon16} & 16 & 46 & 25.5\\ 
 JAPAN & 19 & 62 & 225.3\\
 US24\cite{us24} & 24 & 86 & 755.9\\
 GERMANY\cite{nicolas_vnf} & 50 & 176 & 108000\\
 \hline 
 \end{tabular}
 \caption {Mean run time across networks (in seconds).}
 \label{table:scalability}
\end{table}

Scalability of a solution determines its applicability in real scenarios. So, we show mean run times of our Two-Phase model for networks of different sizes in Table \ref{table:scalability}. Run time is the second phase ($CG+ILP$) execution time. First phase ($SPTG$) execution times were excluded as they were found to be negligible compared to second phase. Note that the Two-Phase model scales well for all networks.

\section{Conclusion}
\label{concl}

We introduce the problem of multiple service chain (SC) mapping with multiple SC {\SCcopies} in presence of highly-populated traffic demands. 
We developed a Two-Phase model, based on a column-generation model along with a Shortest-Path Traffic Grouping (SPTG) heuristic which results in a scalable linear model, thereby solving this complex problem in a relatively small amount of time. Further, we demonstrate that a near-optimal network resource consumption can be achieved with a relatively small number of SC {\SCcopies}, NFV nodes, and VNF replicas for a 100\% populated traffic matrix. This is critical to reduce the network operator's orchestration overhead and capital expenditures. 

\section*{Acknowledgment}
This work was supported by NSF Grant No. CNS-1217978.

\bibliographystyle{IEEEtran}
\bibliography{gupta_nfv}

\begin{thebibliography}{10}
\providecommand{\url}[1]{#1}
\csname url@samestyle\endcsname
\providecommand{\newblock}{\relax}
\providecommand{\bibinfo}[2]{#2}
\providecommand{\BIBentrySTDinterwordspacing}{\spaceskip=0pt\relax}
\providecommand{\BIBentryALTinterwordstretchfactor}{4}
\providecommand{\BIBentryALTinterwordspacing}{\spaceskip=\fontdimen2\font plus
\BIBentryALTinterwordstretchfactor\fontdimen3\font minus
  \fontdimen4\font\relax}
\providecommand{\BIBforeignlanguage}[2]{{%
\expandafter\ifx\csname l@#1\endcsname\relax
\typeout{** WARNING: IEEEtran.bst: No hyphenation pattern has been}%
\typeout{** loaded for the language `#1'. Using the pattern for}%
\typeout{** the default language instead.}%
\else
\language=\csname l@#1\endcsname
\fi
#2}}
\providecommand{\BIBdecl}{\relax}
\BIBdecl

\bibitem{etsi_nfv}
ETSI, ``Network functions virtualisation: Introductory white paper,''
  \url{portal.etsi.org/NFV/NFV_White_Paper.pdf}, 2012.

\bibitem{ietf_sc}
IETF, ``Network service chaining problem statement,''
  \url{https://tools.ietf.org/html/draft-quinn-nsc-problem-statement-00}, 2013.

\bibitem{sc_related}
S.~Mehraghdam, M.~Keller, and H.~Karl, ``Specifying and placing chains of
  virtual network functions,'' in \emph{IEEE 3rd International Conference on
  Cloud Networking (CloudNet)}, Oct 2014, pp. 7--13.

\bibitem{vnf_placement_turck}
H.~Moens and F.~De~Turck, ``{VNF-P: A model for efficient placement of
  virtualized network functions},'' in \emph{10th International Conference on
  Network and Service Management (CNSM)}, Nov. 2014, pp. 418--423.

\bibitem{place_vnf_secci}
B.~Addis, D.~Belabed, M.~Bouet, and S.~Secci, ``Virtual network functions
  placement and routing optimization,''
  \emph{https://hal.inria.fr/hal-01170042/}, 2015.

\bibitem{vnf_placement_barcellos_gaspary}
M.~C. Luizelli, L.~R. Bays, L.~S. Buriol, M.~P. Barcellos, and L.~P. Gaspary,
  ``{Piecing together the NFV provisioning puzzle: Efficient placement and
  chaining of virtual network functions},'' in \emph{IFIP/IEEE Intl. Symp. on
  Int. Netw. Mgmt (IM)}, May 2015, pp. 98--106.

\bibitem{orc_vnf_boutaba}
\BIBentryALTinterwordspacing
M.~Bari, S.~Chowdhury, R.~Ahmed, and R.~Boutaba, ``On orchestrating virtual
  network functions in {NFV},'' \emph{Computing Research Repository}, vol.
  abs/1503.06377, 2015. [Online]. Available:
  \url{http://arxiv.org/abs/1503.06377}
\BIBentrySTDinterwordspacing

\bibitem{sc_detail}
M.~Savi, M.~Tornatore, and G.~Verticale, ``Impact of processing costs on
  service chain placement in network functions virtualization,'' in \emph{IEEE
  Conference on Network Function Virtualization and Software Defined Network
  (NFV-SDN)}, 2015, pp. 191--197.

\bibitem{nicolas_vnf}
\BIBentryALTinterwordspacing
N.~Huin, B.~Jaumard, and F.~Giroire, ``{Optimization of Network Service Chain
  Provisioning}.'' [Online]. Available: \url{https://hal.inria.fr/hal-01476018}
\BIBentrySTDinterwordspacing

\bibitem{zhu_tnsm}
J.~Liu, W.~Lu, F.~Zhou, P.~Lu, and Z.~Zhu, ``{On Dynamic Service Function Chain
  Deployment and Readjustment},'' \emph{IEEE Transactions on Network and
  Service Management}, vol.~PP, no.~99, 2017.

\bibitem{replica_jukan}
\BIBentryALTinterwordspacing
F.~Carpio, W.~Bziuk, and A.~Jukan, ``{Replication of Virtual Network Functions:
  Optimizing Link Utilization and Resource Costs},'' \emph{Computing Research
  Repository (CoRR)}, vol. abs/1702.07151, 2017. [Online]. Available:
  \url{http://arxiv.org/abs/1702.07151}
\BIBentrySTDinterwordspacing

\bibitem{replica_michstate}
\BIBentryALTinterwordspacing
Y.~Jia, C.~Wu, Z.~Li, F.~Le, and A.~X. Liu, ``{Online Scaling of {NFV} Service
  Chains across Geo-distributed Datacenters},'' \emph{Computing Research
  Repository (CoRR)}, vol. abs/1611.08086, 2016. [Online]. Available:
  \url{http://arxiv.org/abs/1611.08086}
\BIBentrySTDinterwordspacing

\bibitem{replica_dcs_cuhk}
X.~Fei, F.~Liu, H.~Xu, and H.~Jin, ``{Towards load-balanced VNF assignment in
  geo-distributed NFV Infrastructure},'' in \emph{IEEE/ACM 25th International
  Symposium on Quality of Service (IWQoS)}, June 2017.

\bibitem{ilp_report}
A.~Gupta, M.~Habib, P.~Chowdhury, M.~Tornatore, and B.~Mukherjee, ``{Joint
  Virtual Network Function Placement and Routing of Traffic in Operator
  Networks},'' \emph{Technical Report, UC Davis}, 2015.

\bibitem{jaumard_colgen_rwa}
B.~Jaumard, C.~Meyer, and B.~Thiongane, ``On column generation formulations for
  the {RWA} problem,'' \emph{Discrete Applied Mathematics}, vol. 157, pp.
  1291--1308, 2009.

\bibitem{jau17RWA}
B.~Jaumard and M.~Daryalal, ``Efficient spectrum utilization in large scale
  {RWA} problems,'' \emph{IEEE/ACM Transactions on Networking}, 2017.

\bibitem{gupta_icc_arxiv}
\BIBentryALTinterwordspacing
A.~Gupta, B.~Jaumard, M.~Tornatore, and B.~Mukherjee, ``{Multiple Service Chain
  Placement and Routing in a Network-enabled Cloud},'' \emph{Computing Research
  Repository (CoRR)}, vol. abs/1611.03197, 2016. [Online]. Available:
  \url{http://arxiv.org/abs/1611.03197}
\BIBentrySTDinterwordspacing

\bibitem{farhan_cost239}
M.~F. Habib, M.~Tornatore, M.~De~Leenheer, F.~Dikbiyik, and B.~Mukherjee,
  ``{Design of disaster-resilient optical datacenter networks},'' \emph{Journal
  of Lightwave Technology}, vol.~30, no.~16, pp. 2563--2573, 2012.

\bibitem{cisco_vnf}
Cisco, ``{Cisco Cloud Services Router 1000V 3.14 Series Data Sheet},''
  \url{http://www.cisco.com/c/en/us/products/collateral/routers/cloud-services-router-1000v-series/datasheet-c78-733443.pdf},
  2015.

\bibitem{skewed_traffic}
R.~Hulsermann, A.~Betker, M.~Jager, S.~Bodamer, M.~Barry, J.~Spath, C.~Gauger,
  and M.~Kohn, ``A set of typical transport network scenarios for network
  modelling,'' \emph{ITG FACHBERICHT}, vol. 182, pp. 65--72, 2004.

\bibitem{eon16}
N.~M. Garcia, P.~P. Monteiro, M.~M. Freire, J.~R. Santos, and P.~Lenkiewicz,
  ``A new architectural approach for optical burst switching networks based on
  a common control channel,'' \emph{Optical Switching and Networking}, vol.~4,
  no.~3, pp. 173--188, 2007.

\bibitem{us24}
S.~Ferdousi, F.~Dikbiyik, M.~F. Habib, M.~Tornatore, and B.~Mukherjee,
  ``Disaster-aware datacenter placement and dynamic content management in cloud
  networks,'' \emph{Journal of Optical Communications and Networking}, vol.~7,
  no.~7, pp. 681--694, 2015.

\end{thebibliography}

\end{document}